\documentclass[11pt]{article}
\usepackage{fullpage}
\usepackage[margin=1in]{geometry}

\usepackage{amsthm}
\usepackage{amsfonts}
\usepackage{amsmath}
\usepackage{morefloats}
\usepackage{authblk}
\usepackage{subfigure}
\usepackage[pdftex]{graphicx} 
\usepackage{url} 

\newcommand{\e}{{\mathbf E}}

\newcommand{\p}[1]{{\mathbf P}\left\{#1\right\}}

\newcommand{\bea}{\begin{eqnarray}}
\newcommand{\eea}{\end{eqnarray}}

\newcommand{\mean}{{\text mean}}

\usepackage{color,soul}

\begin{document}

\title{On the Hyperbolicity of Large-Scale Networks}

\author{W. Sean Kennedy\thanks{Mathematics of Networks Department, Bell Labs, Alcatel-Lucent, Murray Hill, NJ 07974}, Onuttom Narayan\thanks{Department of Physics, University of California, Santa Cruz, CA 95064}, Iraj Saniee$^*$
}
\affil{\texttt{\{kennedy,iis\}@research.bell-labs.com}, \texttt{onarayan@ucsc.edu}}

\maketitle

\begin{abstract}
Through detailed analysis of scores of publicly available data sets 
corresponding 
to a wide range of large-scale networks, from communication and
road networks to various forms of social networks, we explore
a little-studied geometric characteristic of real-life networks, 
namely their hyperbolicity.  
In smooth geometry, hyperbolicity captures the notion of negative curvature;
within the more abstract context of metric spaces, it can be generalized as 
$\delta$-hyperbolicity or negative curvature \emph{in the large}.  This 
generalized definition can be applied to 
graphs, which we explore in this report. We provide strong evidence that
large-scale communication and social networks exhibit this fundamental 
property, and through extensive computations we quantify the degree of 
hyperbolicity of each network in comparison to its diameter.  By contrast,
and as evidence of the validity of the methodology, applying the same 
methods to graphs of road networks shows that they are not hyperbolic, which
is as expected.
Finally, we present practical computational 
means for detection of hyperbolicity and show how the test itself may be scaled
to much larger graphs than those we examined via 
renormalization group methodology. 
Using well-understood mechanisms, we provide evidence through
synthetically generated graphs that hyperbolicity is preserved and 
indeed amplified by renormalization. This allows us to detect hyperbolicity 
in large networks efficiently, through much smaller renormalized versions.
These observations indicate that 
$\delta$-hyperbolicity is a common feature of large-scale networks, 
from IP-layer connectivity to citation, collaboration, co-authorship, 
and friendship graphs. 
We propose that $\delta$-hyperbolicity in 
conjunction with other local characteristics of networks, 
such as the degree distribution and clustering coefficients, provide a more 
complete unifying picture of networks, and helps classify in a 
parsimonious way what is otherwise a bewildering and complex array of 
features and characteristics specific to each natural and man-made 
network. 
\end{abstract}

\pagenumbering{arabic}

\section{Introduction}

Networks arise naturally as an underlying structure in the study of data sets.
These data sets are becoming massive---progressively being collected in larger volumes from the numerous interfaces to our digital life, partly because of the abundance of real-time measurements and inexpensive storage.  
For example, it is now possible to record with unprecedented precision the details of digital messaging, such as the involved parties, the exact time, duration and location, and the medium used.
Extracting useful information from these data sets requires the creation of primitives which can help answer various high level questions.
Perhaps the simplest and most natural primitive is a network or graph structure where high-level questions can be reformulated as algorithmic questions and attacked using various computational methods.
For example, determining which device originated a maximal number of connections is equivalent to determining the degree of a node  in the underlying network, and determining which group of devices had the largest number of  connections between themselves in a specified time period is possible by using the cut value of a group of nodes in the underlying graph.  

A vast body of work has emerged over the past two decades,
dealing with many aspects of the complex networks 
that arise in a variety of data mining settings, such as graphs arising in 
communication, social and biological data sets; for a small sample, see \cite{SNAP,MSW02,FIR02,UKBM11,CGMV09}.  
As in any scientific investigation, the variety and complexity of 
these large and disparate data sets has led to a search for possible 
parsimonious properties underlying their graph primitives.  
Prominent among such properties are the 
(1) small world property~\cite{WS98}, (2) power-law degree distribution~\cite{WS98,BAJ00,FFF09}, 
(3) high clustering coefficients~\cite{WS98}, and (4) existence 
of (small) communities~\cite{DLLM09}.  There have also been several
phenomenological investigations whose aim is to correlate these
properties to each other, such as relating property (1) to a 
combination of properties (2) and (3) \cite{WS98,BAJ00} or (4) to (1)
and indirectly to (2) and (3)
\cite{DLLM09}.  We observe that whereas properties (2) and (3) are
inherently {\em local}, that is, they relate to measures specific to 
nodes and their neighbors, properties (1) and (4) are {\em global} as
they relate to the full graph.  

In this paper, our aim is to highlight 
large-scale geometric features of complex networks as an additional aid
in understanding and classifying large data sets.
Specifically, our goal is to provide evidence through 
computation for another global intrinsic geometric property:
that a large majority of the graphs 
associated with these data sets exhibit hyperbolicity or negative curvature in the large.
This \emph{intrinsic} feature of networks has hitherto not been 
investigated via direct measurements broadly, with the exception of 
1) use of intrinsic hyperbolicity to quantify Internet security~\cite{JL04} and
2) evidence for hyperbolicity in communication networks in the context of 
flow behavior and congestion~\cite{NS11}. However, hyperbolic \emph{embeddings} 
have been
1) studied for efficient geographic routing~\cite{Kle07},
2) leveraged for distance estimation in Internet graphs~\cite{ST08},
3) proposed as an aid for efficient IP inter-domain routing~\cite{BCK09}
and 4) used to approximate distance computation on massive graphs~\cite{SALA11}.
Here we analyze various networks collected by other researchers to 
provide further evidence that (intrinsic) negative curvature exists in 
social and communication networks and more generally it may be meaningfully defined 
and measured on finite graphs, and that even though hyperbolicity is 
distinct from local properties (2) and (3), it can coexist with them.
We have shown earlier~\cite{NS11} that, with few exceptions, hyperbolicity 
implies property (1); thus it can provide an alternative explanation of the small
world property.

We also provide arguments and evidence that, when networks are renormalized 
by combining neighboring nodes into aggregate `supernodes' in a uniform manner, 
the reduced graph is only hyperbolic if
the original network is so, in which case the curvature increases. Thus one can reduce
extremely large networks to smaller renormalized versions, for which the computation of
curvature is a much simpler task. This permits scaling our hyperbolicity test to very large
instances of networks, orders of magnitude larger than current state of
knowledge~\cite{INRIA}.

The rest of this paper is organized as follows. 
In Section \ref{sec:NM}, we summarize some of the standard metrics used in classification  of large-scale networks.  
Section \ref{sec:curvature} gives an introduction to the notion of large-scale curvature and different mechanisms for its computation via a statistical technique for
detecting it~\cite{NS11}.  
The sections that follow summarize some of the key measures of a multiplicity of large graphs as observed and measured by other researchers. 
Specifically, in Section \ref{sec:networks_studied} we discuss over three dozen publicly available and private networks that we use in our study.  
In Section \ref{sec:evidence}, we detail our main observations regarding the intrinsic geometry of these networks.
In Section~\ref{sec:rg} we discuss the renormalization of networks and its effect on hyperbolicity.
Finally in Section \ref{sec:conclude}, we give some concluding remarks.  

\section{Network Measures}
\label{sec:NM}
A principal goal of studying real or man-made networks is identification of their 
key features which help quantify or approximate capacity, robustness, reliability
and other properties.
To this end, several fundamental network features have been identified which
we review in this section.  
These measures fall into two groups: standard measures traditionally used to classify 
large scale networks, and measurements of curvature which are the focus of this work.

\subsection{Standard Measures}

\paragraph{\em Degree Distribution:} The {\em degree} of a node $v$ in a network $G$, denoted $\deg (v)$, is the number of links incident on that node.  
The {\em degree distribution} of the network $G$ for $k = 1, 2, ...$, denoted $P(k)$, is the fraction of nodes in the network that have degree $k$.
Equivalently, if we choose a node $v$ of $G$ uniformly at random, then $P(k)$ is the probably that the degree of $v$ is exactly $k$, i.e., $\p{\deg(v) = k} = P(k)$.
A particularly well-studied distribution in the context of large graphs is the {\em power-law distribution}, that is, when $P(k) \sim k^{-\gamma},$ where $\gamma >1$ is a constant.
The class of networks with such a degree distribution is commonly referred to as  {\em scale-free} networks, and is thought to include many important complex and large-scale networks such as the Internet \cite{FFF09}, the World Wide Web \cite{WS98} and various social networks \cite{N03}.  As an example, 
Figure~7 shows typical nodal degree distributions of social networks
on log-scale charts.  
By contrast, as shown in Figure~8, road networks clearly do not have a 
power-law degree distribution, which is as one would expect.
We remark that several recent studies of various Facebook and other networks have also concluded that their degree distribution is not strictly power-law; 
rather, these distributions are better fitted by gluing together two or more power-law distributions~\cite{GKBM10,UKBM11}.

\paragraph{\em Clustering Coefficient:}  The {\em clustering coefficient} $c_v$ of a node $v$ in a network $G = (V,E)$ is the fraction of pairs of neighbors of $v$ which are adjacent.
Specifically, let $N(v)$ be the neighborhood of $v,$ defined as all the nodes adjacent to $v.$ Let $E(N(v))$ be the number of edges with both 
endpoints contained in $N(v).$ Then
$$c_v = \frac{|E(N(v))|}{{N(v) \choose 2}} = \frac{2|E(N(v))|}{|N(v)|(|N(v)| - 1)}.$$
Intuitively, $c_v$ is the fraction of all triplets consisting of a vertex $v$ and 
two of its neighbors that are triangles. It quantifies how close $v \cup N(v)$ is to being a clique or, in other words, a tightly knit community.  
For each network we record the mean and median clustering coefficient over the whole node set.  
We note that in all cases the minimum clustering coefficient  is zero and the maximum is one.  

\paragraph{\em Characteristic Path Length and Diameter:}  For any connected graph $G$, the {\em characteristic path length}, denoted $\e d(x,y)$, is the expected distance between any two different vertices chosen uniformly at random.   This distance scales like the {\em diameter} of the graph, the 
maximum distance between any pair of vertices.
In cases where the network is so large that it prohibits computing all pairwise distances, we compute instead $\e^\star d(x,y)$ equal to the average $d(x,y)$ over a fixed percentage of all node pairs in the network chosen uniformly at random \cite{WS98}.
There has been much discussion in the literature about the ubiquitousness
of small characteristic path lengths in the majority of natural and man-made networks;
a small characteristic path length means that this length scales like $O(\log n)$ or 
$O(\log \log n)$ for a network of $n$ nodes,
rather than $O(n^\alpha), \alpha \leq 1$.  
We note that, Watts and Strogatz~\cite{WS98} define a {\em small-world network} as one with both a large clustering coefficient and small characteristic path length.  

\subsection{Measures of Curvature}
\label{sec:curvature}
The fundamental notion of curvature comes from differential geometry 
where it is traditionally defined for ``smooth" mathematical structures
such as manifolds.  It turns out that curvature may be defined
analogously and consistently for a much wider class of objects within 
the context of metric spaces, which naturally includes infinite graphs with 
appropriate link weights.  This remarkable generalization is more recent 
(see Refs.~\cite{Gromov87,BH99,dlH00}) and
its further development is under active investigation~\cite{BOW90}.
Early applications of this theory includes \cite{JL02,Loh03,JL04,AJL11}.
   
In~\cite{BJL08}, it was observed that the notion of curvature as defined 
for metric spaces can be extended to 
graphs that are large but finite, creating
a novel feature which has not been quantified or
investigated for real networks before.  In addition, ~\cite{NS11} proposed an
algorithmic means to characterize and quantify  
hyperbolicity in networks, particularly large-scale networks.
Here we summarize some of the basic concepts of this toolbox.

Assigning a non-negative weight to each edge of a graph, e.g. a unit weight for all
the edges, associates a {\it length} to each 
path in a (finite or infinite) graph,  
defined as the sum of the weights of all edges traversed by the path.  
For instance, in a communication network, the length of a path may be  
the (routing) cost associated with the transfer of each
packet that traverses this path.
The {\it distance}, written as $d(A,B)$, between two vertices $A$ and $B$ in the 
graph is the  
minimum of the lengths of all paths between them. A path of  
shortest length between the end points is also called a {\bf geodesic}.
A metric space is a set of points for which any two points
have an associated shortest distance between them
that satisfy the three basic metric (including triangle) inequalities.  
Thus a graph with a non-negative edge weight may be viewed as a metric space.  
A geodesic (or path) metric space
is a metric space in which the shortest distance between any two points is achieved by
at least one path between the points.

There are many equivalent definitions of negative curvature for an
infinite graph. One of the more intuitive ones is as follows:

\paragraph {\em Slim Triangle Condition:}  A geodesic metric space is said to
have the {\em $\delta$-slim triangle property} or be {\em $\delta$-hyperbolic} if 
there exists a  $\delta \geq 0$ such that, for any three points $A, B, C$ of the  
space connected to each other by geodesics (ties are
broken arbitrarily) $AB$, $BC$, $CA$,
it is the case that the union of the $\delta$-neighborhoods
of any two pairs of geodesic triangle sides, say
$N_{\delta}(AB) \cup N_{\delta}(BC)$, includes the 3rd side, $CA$. 

\paragraph {}
When $G$ is a tree, then one can take $\delta =0$.
The motivation underpinning the above condition is that the standard  
comparison theorems in differential geometry show that a simply  
connected manifold whose curvature is negative and bounded away from  
0 satisfies the same condition, where $\delta$ depends on the upper
bound on the Gaussian curvature. Conversely, this condition fails for a  
manifold of curvature $\geq0$ such as the standard Euclidean space.
It has long been known that the curvature of a 
smooth manifold has a big impact on its large-scale properties, 
and similarly and quite remarkably many important properties of metric 
spaces (and thus graphs) may be traced back to the above Slim  
Triangle Condition (or any of its equivalent forms).  

\begin{figure}[htb]
\begin{center}
\includegraphics[width = 3in]{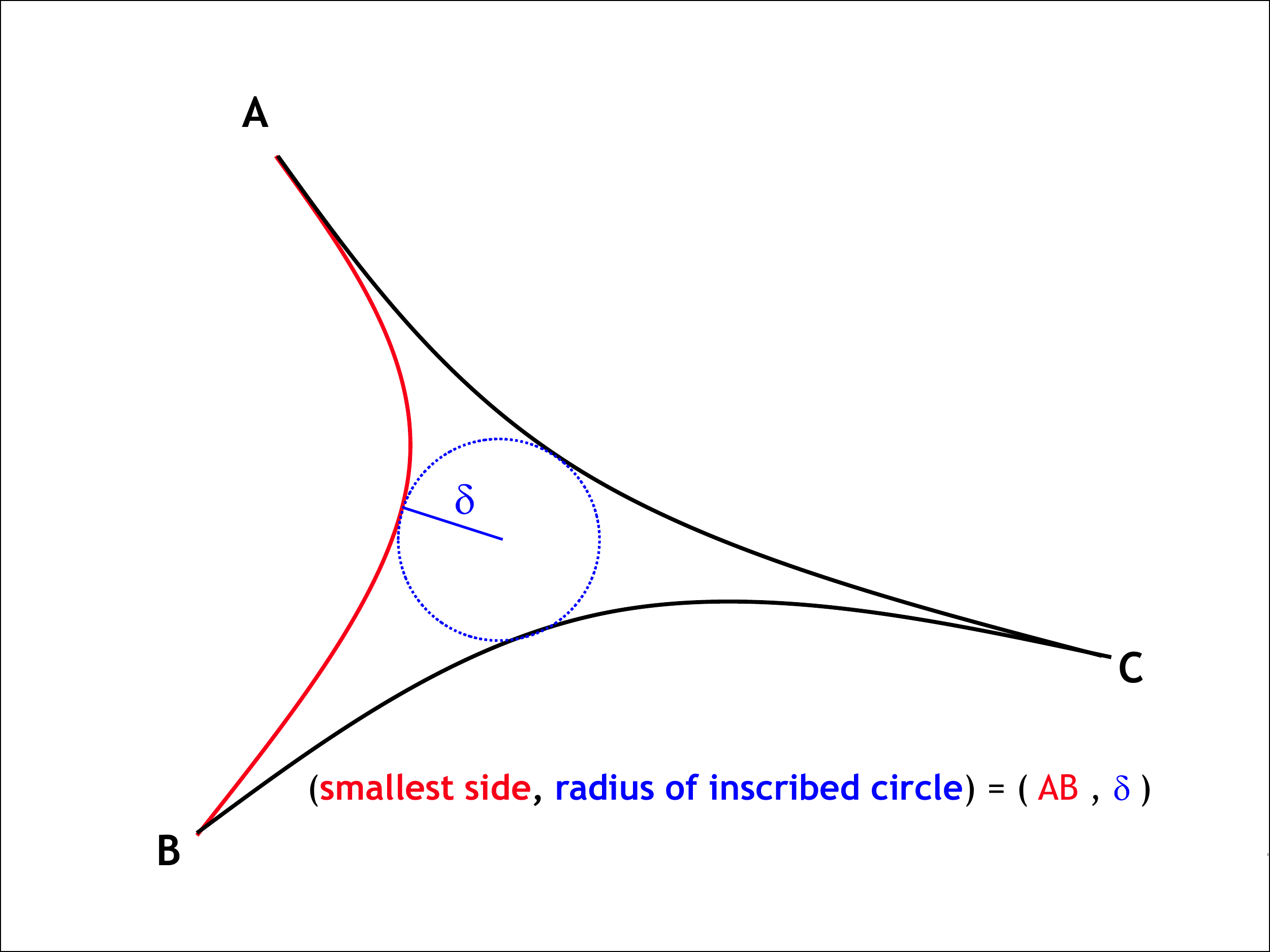}
\caption{A visualization of the thin triangle condition for
$\delta$-hyperbolic spaces.
}
\label{fig:thinTriangle}
\end{center}
\end{figure}

In practice and
for computational ease, we use~\cite{NS11} an equivalent definition of a
\emph{thin triangle}, see \cite{BH99}, as illustrated in 
Figure~\ref{fig:thinTriangle}:  
for any three
nodes $(ABC)$, the geodesic paths $g_{AB}, g_{BC}$ and $g_{CA}$ of lengths
$d_{AB}, d_{BC}$ and $d_{CA}$ are constructed.  A fourth node $m$
is chosen arbitrarily, and the shortest distance between $m$ and all
the nodes on $(AB)$ is defined as $d(m;AB).$ 
Then if
\begin{equation}
\max_{(ABC)}\min_m \max[d(m;AB), d(m; BC), d(m; CA)]
\label{eq:gromov_def}
\end{equation}
is finite, a (possibly infinite) graph is said to have negative or
hyperbolic curvature.
For reasons that will become apparent, we will say such a graph satisfies the {\em $\delta$ 3-point condition} or simply the {\em $\delta$-3PC}.
We note that the computed
$\delta$'s based on the slim and thin triangle conditions are not exactly the same, but 
they only differ slightly and more significantly, they converge and diverge together.
One is typically interested in the smallest such $\delta \geq 0$ if
and when it exists.  

Naturally, any finite graph $G$ of diameter $D$ must satisfy
$\delta(G) \leq D$. Thus for negative curvature to be meaningfully defined for 
finite graphs, the definition given above has to be extended.
Typically, for a family of finite graphs indexed
by $n$ (number of nodes) one looks for $\delta/D$ to be small and asymptotically
zero to call the family $\delta$-hyperbolic~\cite{BJL08}.  
To see how the said toolbox can be applied in a network setting,
we introduce the concept of the ``probability distribution curvature plot''
of a network (introduced in \cite{NS11}):
for every triangle $\Delta = (ABC)$ we plot $\delta_\Delta$
vs $L_\Delta$ where
\begin{eqnarray}
\delta_\Delta &=& \min_m \max[d(m;AB), d(m; BC), d(m; CA)]\nonumber\\
L_\Delta &=& \min [d_{AB}, d_{BC}, d_{CA}].
\label{eq:finitegraph}
\end{eqnarray}
(See Figure 1.)   This yields $P_L(\delta),$ the probability distribution for $\delta$
at fixed $L.$ If the peak of $P_L(\delta)$ is at
$\delta=\delta_p(L),$ the network is flat (negatively curved)
if $\delta_p(L)$ increases linearly (sublinearly) with $L$
\footnote{By choosing $m$ to be on the sides of the triangle, it is easy to verify
that $\delta_\Delta \leq L_\Delta/2,$ so that positive curvature cannot
be seen with this test or Eq.(\ref{eq:gromov_def})}. 
Since we use the peak of the distribution instead
of the maximum as in Eq.(\ref{eq:gromov_def}), statistical sampling of
triangles is sufficient.

\begin{figure}
\begin{center}
\subfigure[Network 7018 (AT\&T network): 
10152~nodes, 4319 links, diameter 12.] {
\label{subfig1}
~
\includegraphics[width = 2.75in]{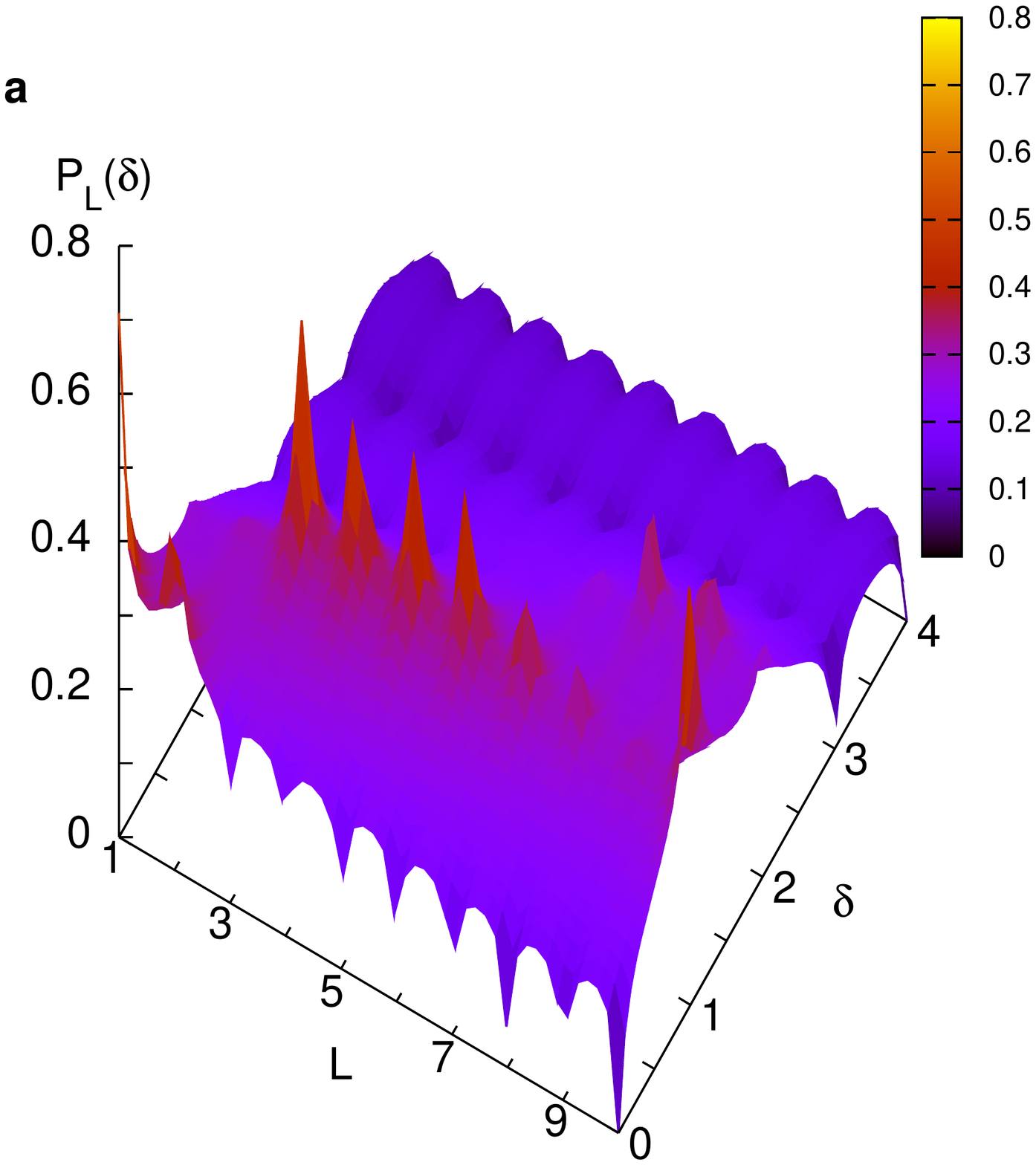}
~}
\hspace{1cm}
\subfigure[Triangular~lattice~(flat): 469~nodes, 1260 links, diameter 23.] {
\label{subfig2}
~
\includegraphics[width = 2.75in]{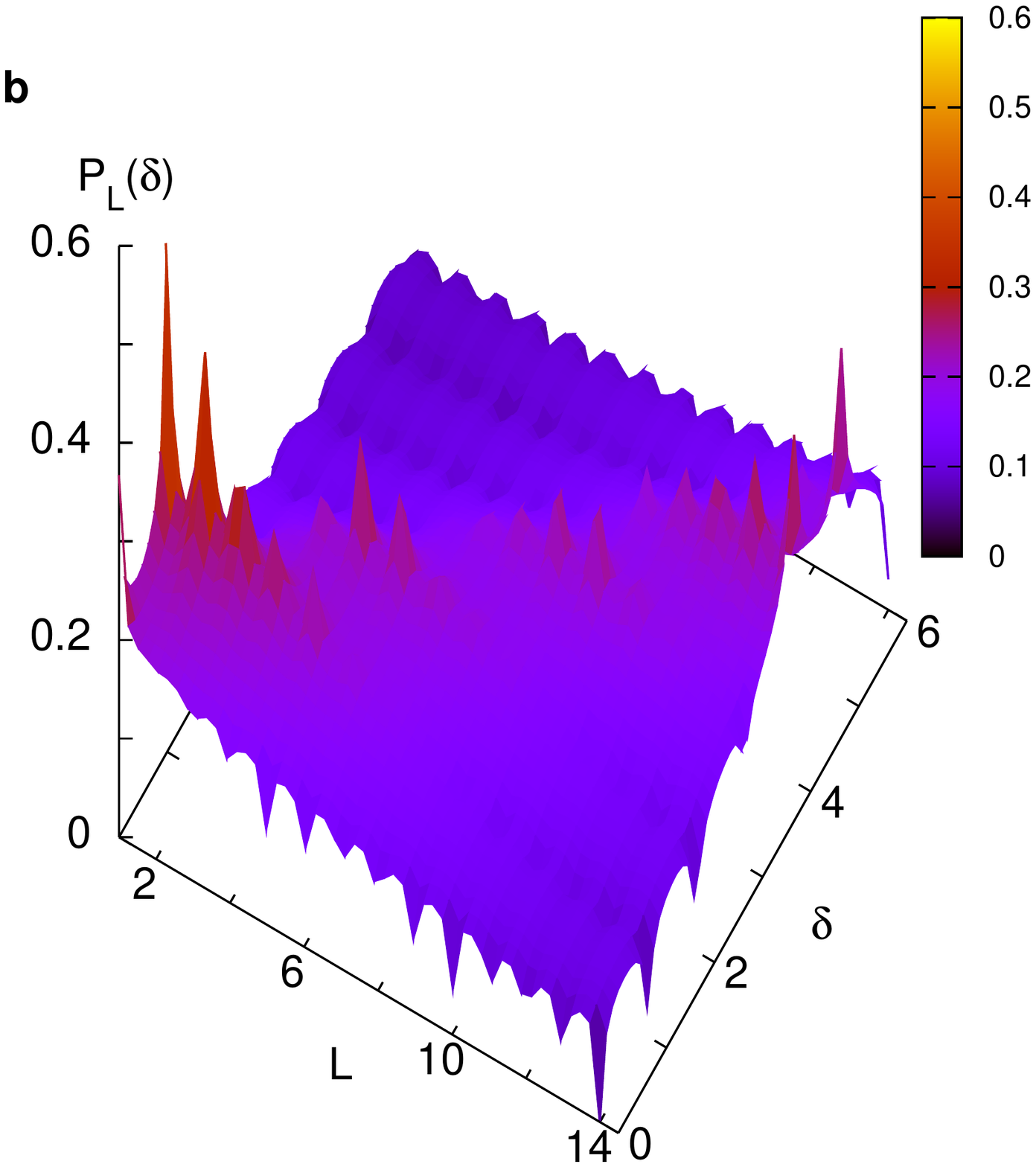}
~}
\end{center}
\caption{Probability Distribution Curvature plots~\cite{NS11} (arXiv version):
Probability $P_L(\delta)$ for randomly chosen triangles whose shortest side is $L$ to have a given $\delta$ as defined in Eq.(\ref{eq:finitegraph}) for two networks.
The quantities $\delta$ and $L$ are restricted to integers, and a smooth plot is determined by interpolation.
Despite the smaller number of nodes in Fig.~\ref{subfig2}, the range of $L$ is similar to that in Fig.~\ref{subfig1} because the triangular lattice is not a 
small world graph.}
\label{fig:gromovplot}
\end{figure}
Figure~\ref{fig:gromovplot} shows the probability distribution 
curvature plot for network
7018 (AT\&T) from the Rocketfuel database~\cite{Rocketfuel}.
The metric used is the 
`hop metric', where each edge of the graph has unit length.
This is a common metric that best illustrates the geometrical
properties of the graph, including the `small world' property~\cite{WS98}.
The networks in this database are at the IP layer and describe the IP
port to IP port connectivity of a network. 
A sharp ridge is seen along the curve
$\delta_p(L).$ The ridge is a straight line through the origin for
the triangular lattice 
but bends over parallel to the $L$-axis for the 7018 network ($P_L(\delta)$
is {\it zero} for $\delta > 3$ for all $L$, though the diameter of the
network is 12).  For all the networks in the database, we have verified
that the measured $\delta$'s do not exceed 3, even though the network
diameters range from  12 to 14 (with the exception of 4755/VSNL whose
diameter is 6, but whose ratio diameter/$\delta$ is even bigger, 6).
The ratio of 3/12 or 25\% is comfortably within the theoretical bound
for scaled hyperbolic graphs~\cite{BJL08}.

\begin{figure}
\begin{center}
\includegraphics[angle=0,width=3.5in]{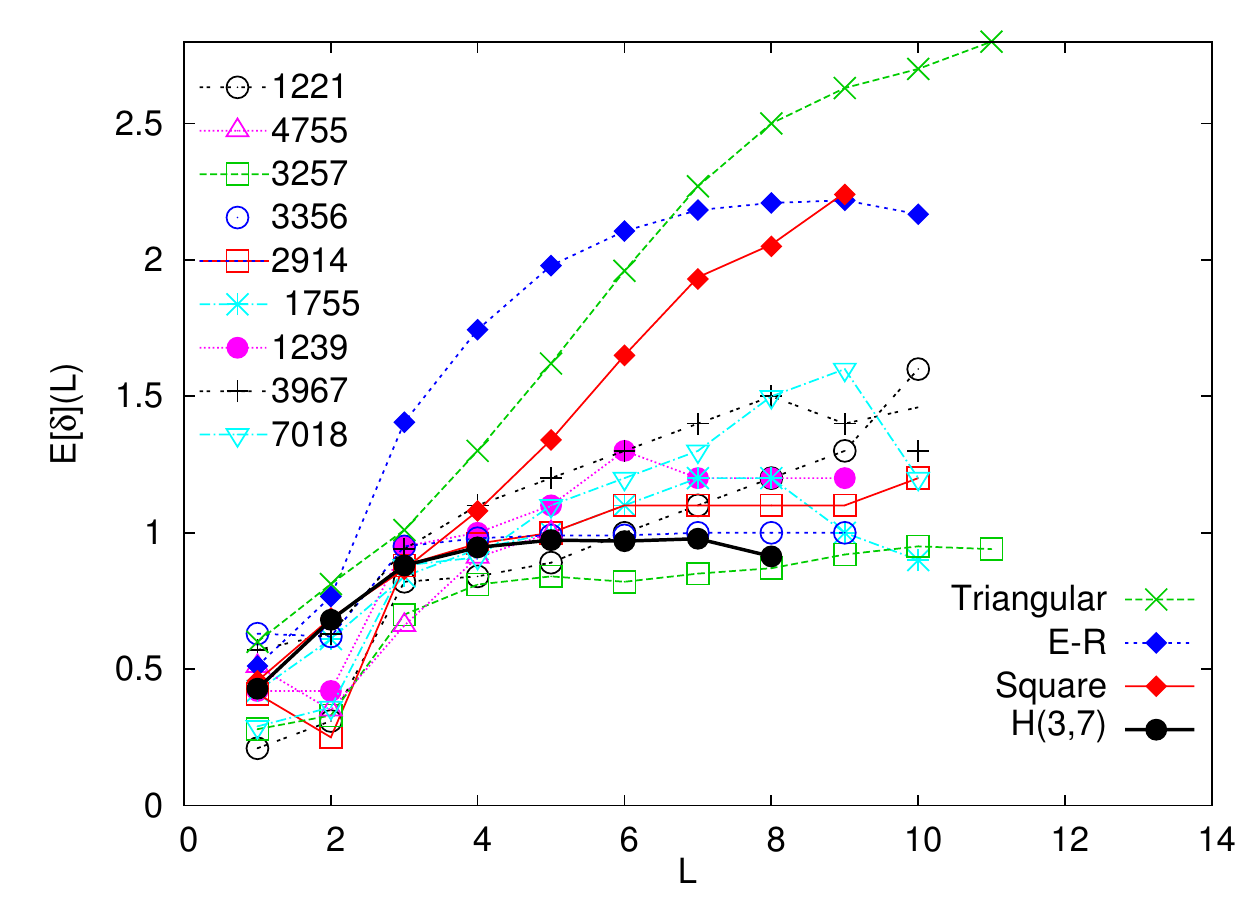}
\caption{3-Point Curvature Plot~\cite{NS11}: The average triangle thickness 
$E[\delta](L)$ as a function of $L$ (the smallest side of the triangle) 
for the 9 IP layer RocketFuel networks, two flat grids 
(the triangular lattice with diameter 29 and the square 
lattice with diameter 154), and an Erd\"os Renyi graph.}
\label{fig:gromovgraph}
\end{center}
\end{figure}

As another manifestation of curvature, Figure~\ref{fig:gromovgraph}~\cite{NS11} 
shows the average $\delta$ for each $L,$ $E[\delta](L),$ for all ten
networks in the Rocketfuel database. The plots 
saturate for relatively small $L.$ The figure also shows $E[\delta](L)$
for the triangular and square lattices (which are obviously not hyperbolic) and the 
hyperbolic grid $\mathbb{H}_{3,7}.$
The qualitative difference between the highly flattened plots for the Rocketfuel 
networks
and the approximately linear plots for the lattices is evident. 

\begin{figure}[h]
\begin{center}
\includegraphics[angle= 0, width =3in]{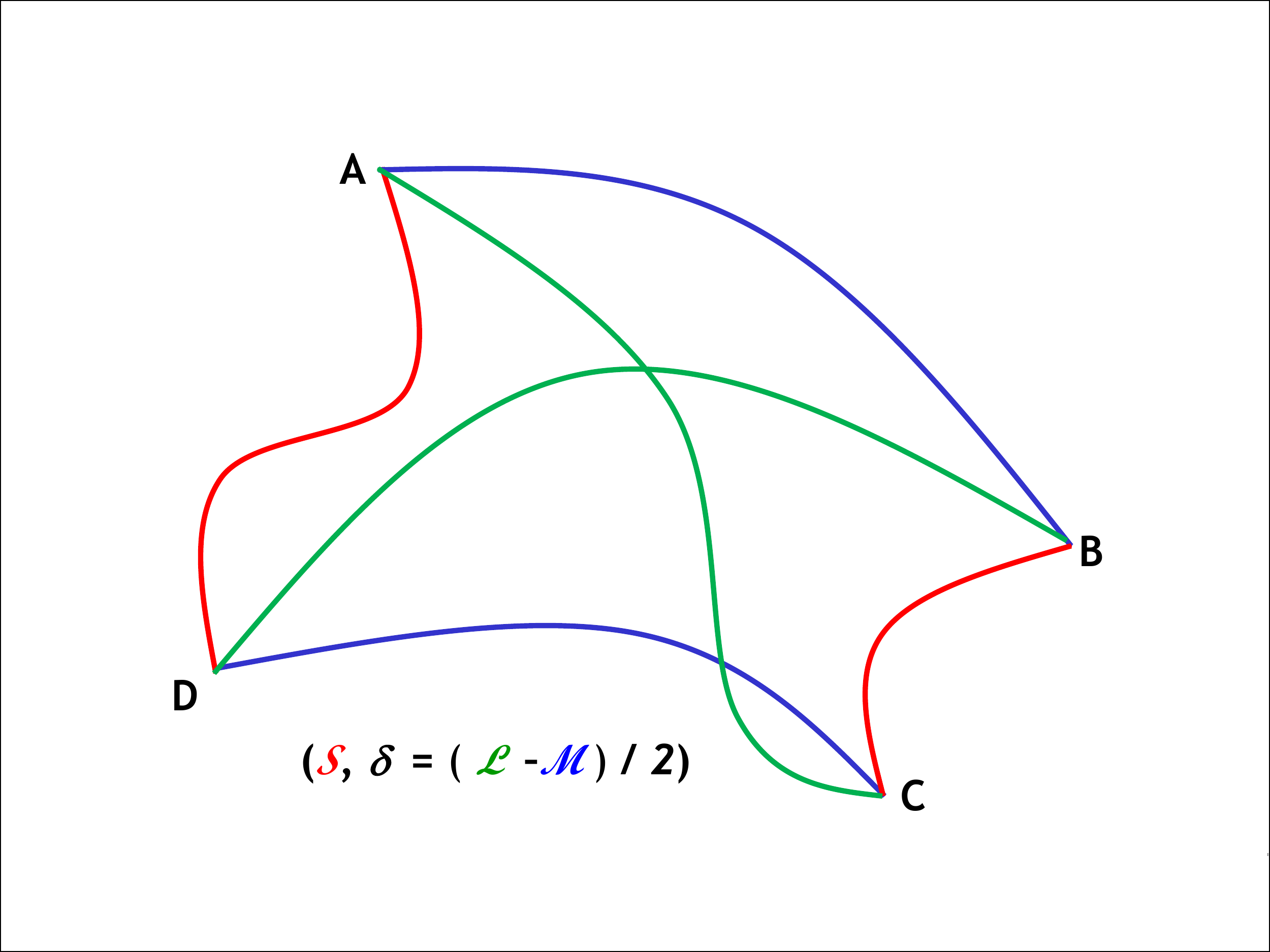}
\caption{Schematic representation of the 4-point condition.}
\label{fig:4pc}
\end{center}
\end{figure}

\paragraph{\em Four Point Condition:} An alternative algorithm for checking for 
hyperbolicity is to use the {\em 4-point condition} which is defined as follows.
Consider an arbitrary set of four points $A,B,C,D$ and the three sums of the opposite 
sides $d(A,B)+d(C,D)$, $d(A,C)+d(B,D)$ and $d(A,D)+d(B,C)$.
We can assume, by relabelling if necessary, that $A,B,C,D$ satisfy 
$$d(A,B)+d(C,D) \le d(A,C)+d(B,D) \le d(A,D)+d(B,C),$$ 
and so we find it convenient to let the largest of these ${\cal L} = d(A,D)+d(B,C)$, 
the middle ${\cal M} = d(A,C)+d(B,D)$, and smallest ${\cal S} = d(A,B)+d(C,D)$, see 
Figure~\ref{fig:4pc} for an illustration. 
For fixed constant $\delta > 0$, $A,B,C,D$ satisfy the {\em $\delta$ 4-point 
condition}, or $\delta${\em-4PC}, if 
$$\frac{{\cal L} - {\cal M}}{2} \le \delta.$$
We are typically interested in the smallest $\delta$ such that every set of four points 
in the graph satisfies the $\delta$-4PC. 

The 4-point condition is intimately related to the 3-point condition.  
Indeed, it is not hard to show that if every triple satisfies the 3PC 
for at most $\delta$ then any set of four points satisfies the $2\delta$-4PC, see for 
example \cite{BH99}.
Conversely, if every quadruple satisfies the $\delta$-4PC, then any set of three points 
satisfies the 3PC for at most $6\delta$.
As shown by Buneman~\cite{Buneman74}, if $G$ is a tree then $G$ satisfies the $0$-4PC 
for all quadruples of nodes in $G$.  
More generally, a graph  $G$ satisfies the $0$-4PC precisely if
every 2-connected subgraph of $G$ is a clique~\cite{How79}.  Note
that in this case we have ${\cal L} = {\cal M}$ for all quadruples.
For a general graph, for any four points $A,B,C,D$ ordered as above, the triangle 
equality implies that
$$\frac{{\cal L}-{\cal M}}{2} = \frac{d(A,D)+d(B,C) - d(A,C) - d(B,D)}{2} \le \min \{d(A,B),d(C,D)\}.$$
So, setting $S_{min} =\min \{d(A,B),d(C,D)\}$ we have the natural upper bound $\frac{{\cal L}-{\cal M}}{2}  \le S_{min}$.
By plotting $\frac{{\cal L}-{\cal M}}{2}$ versus $S_{min}$ we know the worst case growth of the function.  
This is how we depict the 4-point curvature plots.

\begin{figure}[b]
\begin{center}
\subfigure[Standard hyperbolic grid $\mathbb{H}_{3,7}$ with $r$ levels ($2^r$ nodes).] {
\label{subfig3}
~
\includegraphics[height = 1.7in]{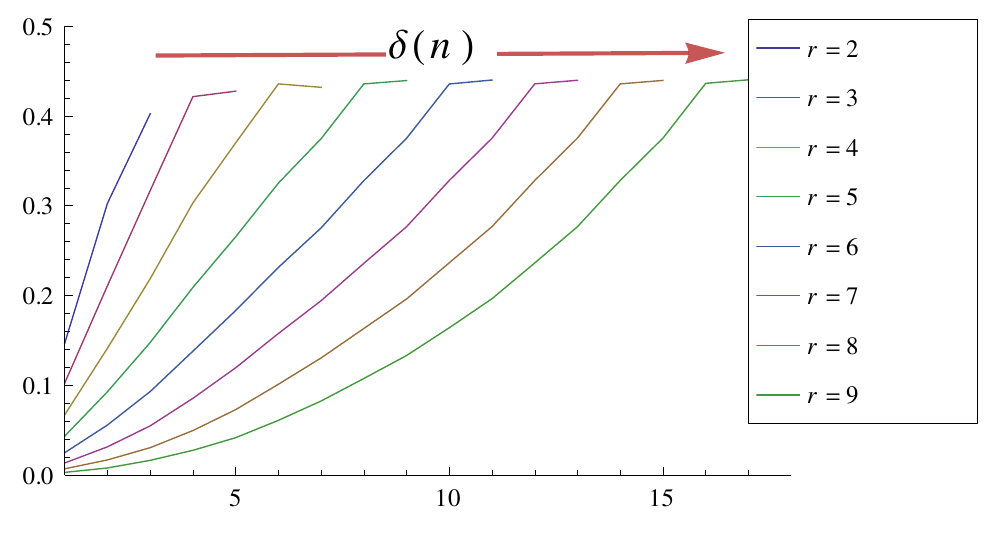}\label{fig:H37}
}
\subfigure[Two-dimensional Euclidean grid] 
{
\label{subfig3-2}
~
\includegraphics[height= 1.7in]{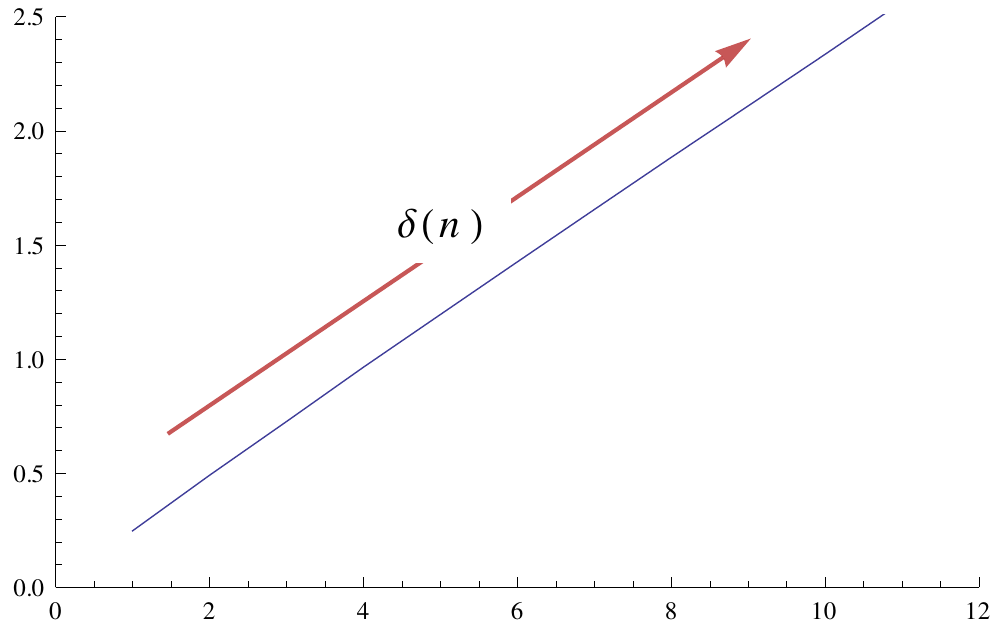}\label{fig:squareGrid}
}

\caption{Curvature plots using the 4-point condition.}

\end{center}
\end{figure}

As was argued for both the 3-point curvature plot (3P-CP) and the 4-point curvature 
plots (4P-CP), if the resulting curve saturates at a small value (compared to graph 
diameter) then this is evidence of hyperbolicity, as shown for two prototypical 
hyperbolic
and non-hyperbolic graphs $\mathbb{H}_{3,7}$ and $\mathbb{Z}^2$, see Figures \ref{fig:H37} and \ref{fig:squareGrid}.
The key advantage of the 4P-CP compared to the 3P-CP is that the former only 
requires 
knowledge of the distance between pairs of points whereas the latter requires 
construction of the actual path. 


\subsection{Computation of Hyperbolicity}
\begin{figure}
\begin{center}
\includegraphics[angle= 0, width =3.5in]{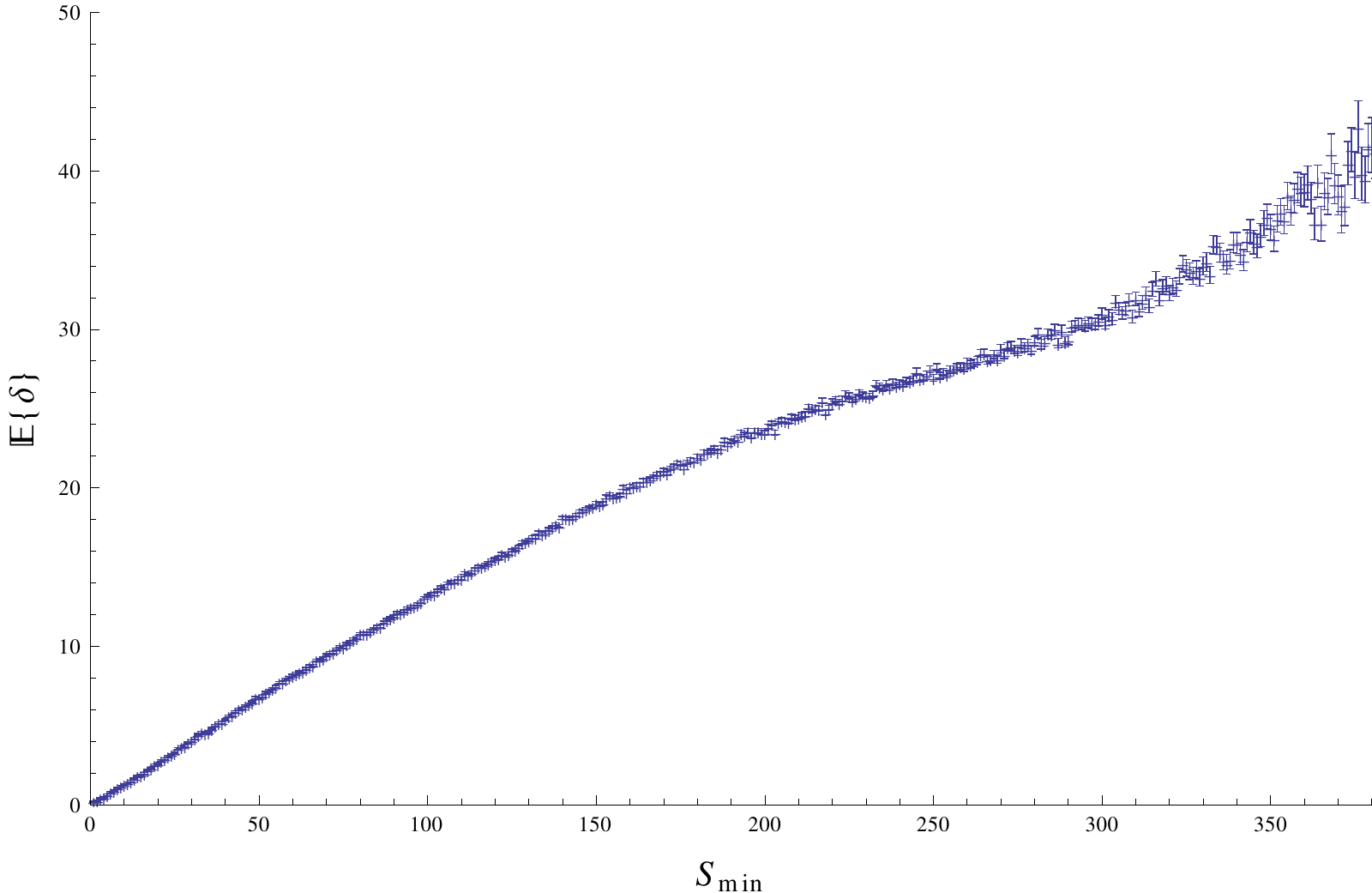}
\caption{A plot of the standard error for the mean in our calculation of the four-point curvature plot for the roadNet-PA network. The error bars are displayed as vertical line segments and are barely visible.}
\label{fig:StdError}
\end{center}
\end{figure}

In this section, we discuss some of the statistical aspects of curvature plots,
in particular, the number of quadruples we need to sample in order to achieve a 
sufficiently high degree of confidence for interpreting the curvature plot.  
For this purpose we use the {\em standard error of the mean} which  
is the standard deviation of the estimate of a population mean given by the 
sampled mean. Formally, we define the standard error of a sample size $s$ as
the standard deviation of the sample divided by $\sqrt{s}$.  
Figure \ref{fig:StdError} gives a bar plot of the standard error for the roadNet-PA Network 
curvature plot (see Section \ref{sec:networks_studied} for details).
Here, in addition to plotting the sampled mean value of $\delta$ for each fixed $S_{min}$,
as our curvature plot requires, we give error bars showing the standard error of each 
point, which is the mean $\delta$. As in this example, for all our curvature plots we sample
quadruples until the standard error at each $S_{min}$ is negligible.

%

\section{Networks Studied}
\label{sec:networks_studied}
We study over three dozen publicly available and private data sets corresponding to a range of large-scale networks.
In this section, we describe these networks.
Overall these networks fall into seven categories, which we will discuss separately.
These categories are social, signed, IP layer, peer-to-peer, collaboration, web, and road networks.
Roughly half of these networks have been well-studied and are available as part of the Stanford Large Network Dataset Collection (SNAP)~\cite{SNAP}.
The IP layer networks are from the (also) well-studied RocketFuel data set~\cite{Rocketfuel}.
For all of these networks, we have retained the original SNAP and Rocketfuel nomenclature for ease of reference and comparison purposes.  
Finally, we include a set of four networks extracted from the social network Facebook.
In all cases the networks we study are undirected, simple and unweighted graphs; in particular, in those cases that the original is directed we study the underlying undirected graph.
A detailed description of these networks now follows.  

\paragraph{Social Networks:}
We study five social networks in total. 
The key measures of these networks can be found in 
Table 1.a and their degree distributions are plotted in 
Figure~7.
Four of these are Facebook networks, the popular online networking site.
Here each node represents a unique user of the social network and two nodes are connected with a link if the two users are ``friends''.  
These are \texttt{sn-Small}, \texttt{sn-Medium} and \texttt{sn-Large} which are three anonymized Facebook datasets from~\cite{PSWZ12} and \texttt{sn-Facebook} publicly available from \cite{CGMV09}.
The final social network we study is the wiki-Vote network from the SNAP data set~\cite{HKL10}, which is built from Wikipedia.com the free online encyclopedia.
To understand this graph, observe that Wikipedia administrators are users who have been promoted through a Wikipedia contributors community voting process.
The network studied has a node for each such user where two users are connected if one user voted on the other user's promotion in all promotion votes between the inception of Wikipedia through January 2008. 
\begin{figure}
\label{degSN-new1}
\begin{center}
\subfigure[\texttt{sn-Small} Network]{
\includegraphics[width=7.5cm]{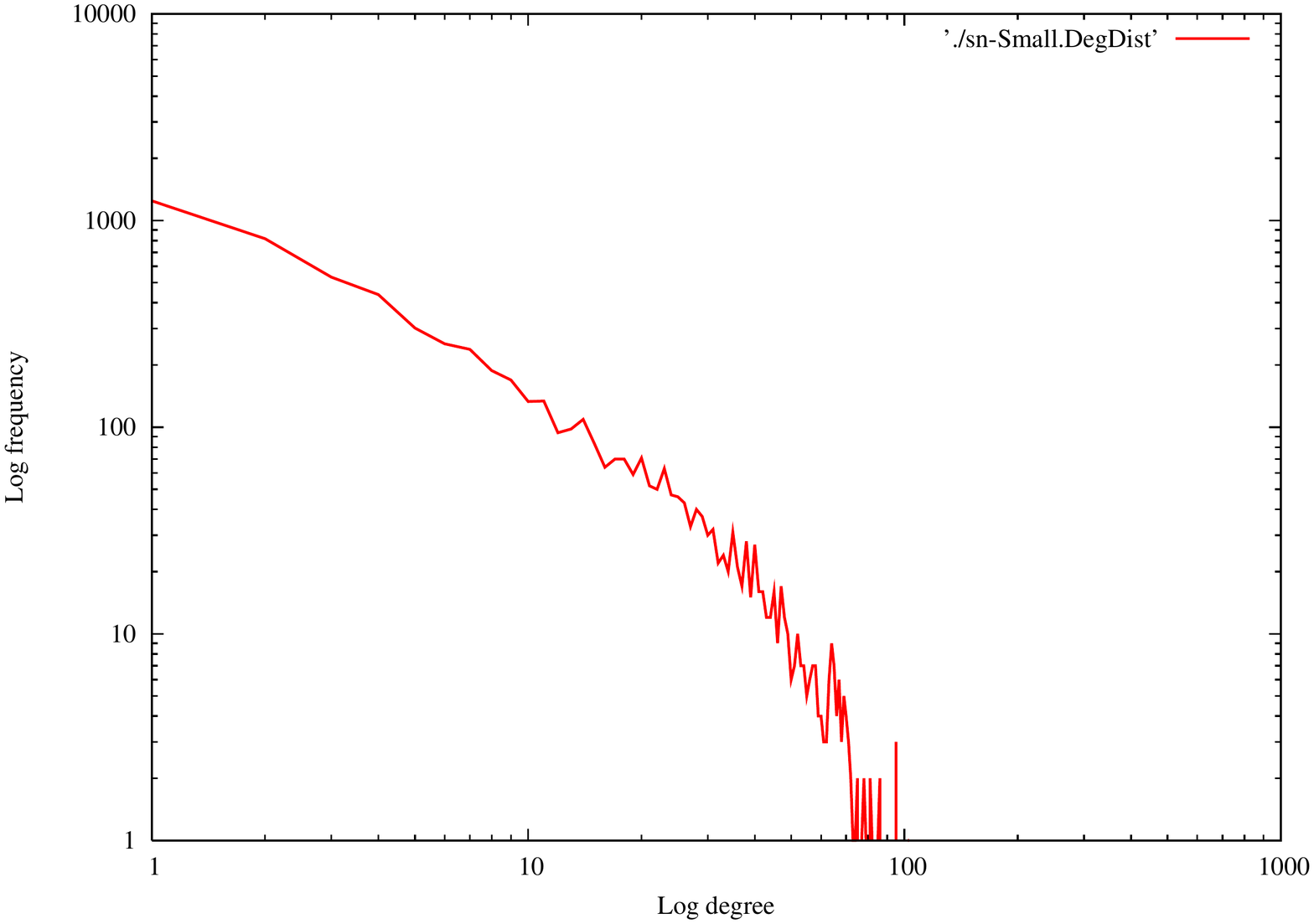}
}
\subfigure[\texttt{sn-Medium} Network]{
\includegraphics[width=7.5cm]{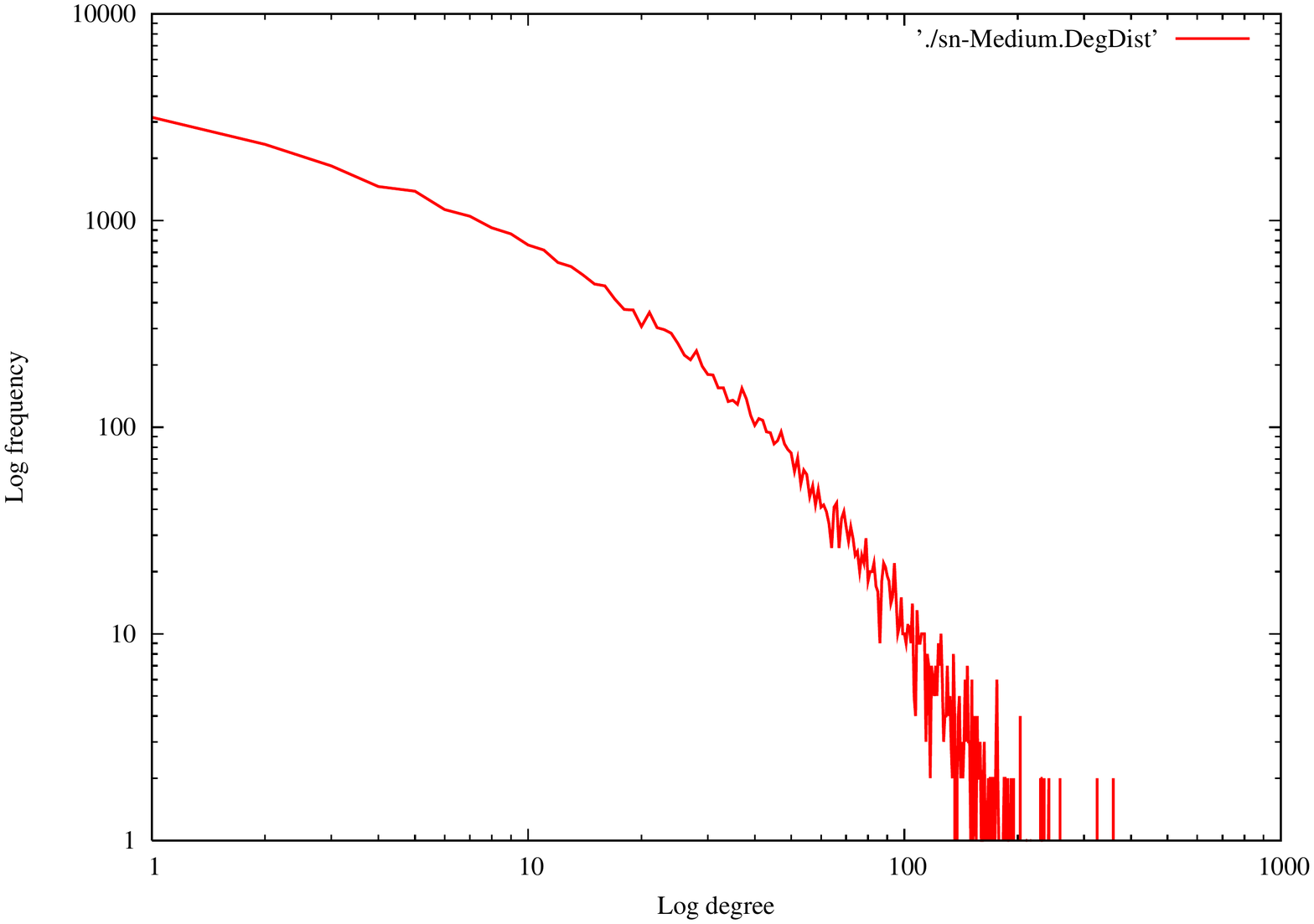}
}
\end{center}
\caption{Degree distributions of two social networks on a log-log plot, indicating
power-law distributions at least in the larger degree regime.}
\end{figure}

\paragraph{Signed Networks:}

We study the underlying undirected graph of six signed networks, which are social networks where users can evaluate other users positively or negatively, for example by ``liking'' or ``disliking'' them.
The network \texttt{soc-sign-Epinions} is derived from the consumer review site Epinions.com.
Here nodes correspond to users and two nodes are adjacent if  corresponding customers ``trust'' each other.  
The five networks \texttt{soc-Slashdot0811}, \texttt{Slashdot081106}, \texttt{Slashdot090216}, \texttt{Slashdot090221}, and \texttt{soc-Slashdot0922}, collected between 2008 and 2009, are built from the technology-related news website Slashdot.com.
Inspired by the Slashdot Zoo feature, users can tag other users as ``friends'' or ``foes''.
These networks are then built with nodes corresponding to users where two nodes are adjacent if one of the corresponding users has tagged the other.  
The standard measures of these networks can be found in Table 1.b.
Further information can be found in \cite{HKL10}

\paragraph{Rocketfuel:}
Nine IP-layer networks as mapped out in 2002 by the 
Rocketfuel ISP topology mapping engine are also studied~\cite{MSW02,Rocketfuel}.
Here nodes correspond to IP ports on routers and links to 
the wires interconnecting them.  
The nine Rocketfuel networks are \texttt{1221 Telstra} (Australia), 
\texttt{1239 Sprintlink} (US), \texttt{1755 Ebone} (Europe), \texttt{2914 Verio} (US), 
\texttt{3257 Tiscali} (Europe), \texttt{3356 Level3} (US), \texttt{3967 Exodus} (US), 
\texttt{4755 VSNL} (India)
and \texttt{7018 AT\&T} (US).
As discussed above, the hyperbolicity of these networks was previously studied in \cite{NS11}.
The key measures of these networks can be found in Table 1.c.

\paragraph{Peer-to-peer Networks:}
From the SNAP data set we study nine internet peer-to-peer networks built from a sequence of nine snapshots of the Gnutella file sharing network. Collected in August of 2002, here nodes correspond to Gnutella hosts and edges represent connections between the respective hosts.
The key measures of these networks can be found in Figure 1.d.
Further information can be found in \cite{FIR02,FKL07}.

\paragraph{Collaboration Networks:}
We study five collaboration networks from the SNAP data set.  
These are built from the e-print service arXiv.org for different scientific research communities who use this service.  
Nodes in these networks correspond to authors of papers, and nodes are connected by an edge whenever the corresponding authors co-authored an article.
In all cases, the data was collected for articles published on the arXiv in between January 1993 to April 2003 and covers nearly the complete history of publications on the arXiv up until April 2003.
The five research communities are astrophysics (\texttt{CA-AstroPh}), Condensed Matter Physics (\texttt{CA-CondMat}),   General Relativity and Quantum Cosmology (\texttt{CA-GrQc}), High Energy Physics---Phenomenology (\texttt{CA-HepPh}), and High Energy Physics---Theory (\texttt{CA-HepTh}).
The key measures of these networks can be found in Table 2.a.
Further information can be found in \cite{FKL07}

\paragraph{Web Networks:}
We study three web networks.
Here nodes correspond to web pages and two nodes are adjacent if the web page corresponding to at least one of them has a hyperlink to the other.
The network \texttt{web-Google} is from Google and was release in 2002.
The networks \texttt{web-BerkStan} corresponds to the berkeley.edu and stanford.edu domains together and the \texttt{web-Stanford} network corresponds to the stanford.edu domain. 
The key measures of these networks can be found in Table 2.b.
Further information can be found in \cite{DLLM09}.

\paragraph{Road Networks:}
The largest networks we study, these three road networks correspond to the road networks of the US states of California, Pennsylvania and Texas.
Here nodes correspond to intersections and other endpoints, for example, dead ends, and edges are the road segments connecting them.
The key measures of these networks can be found in Table 2.c. and the degree distribution for the Texas road network, a non-scale free network, 
is plotted in 
Figure~8. Further information can be found in \cite{SNAP,DLLM09}.

\begin{figure}
\label{degSN-new2}
\begin{center}
\includegraphics[width=7.5cm]{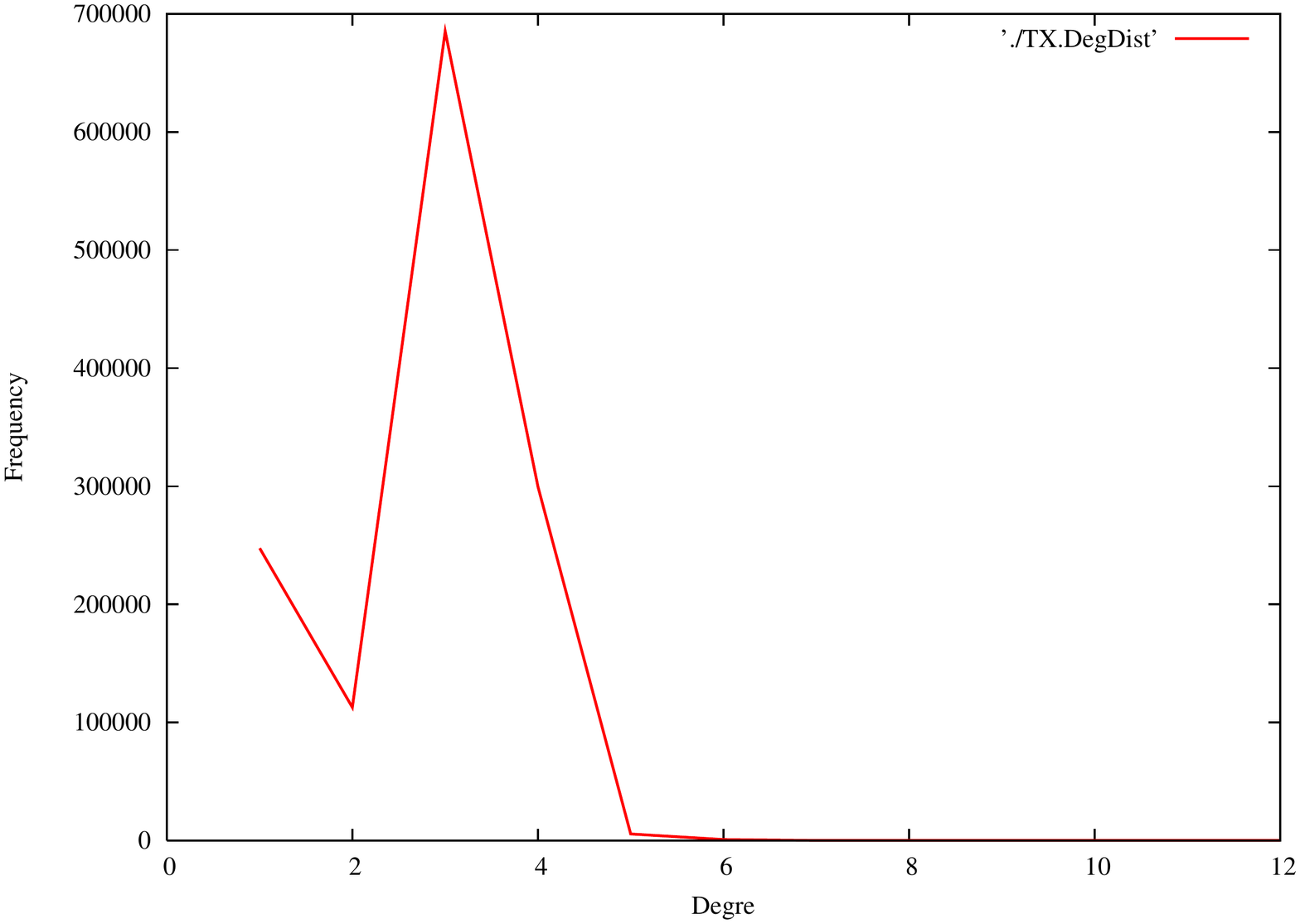}
\includegraphics[width=7.5cm]{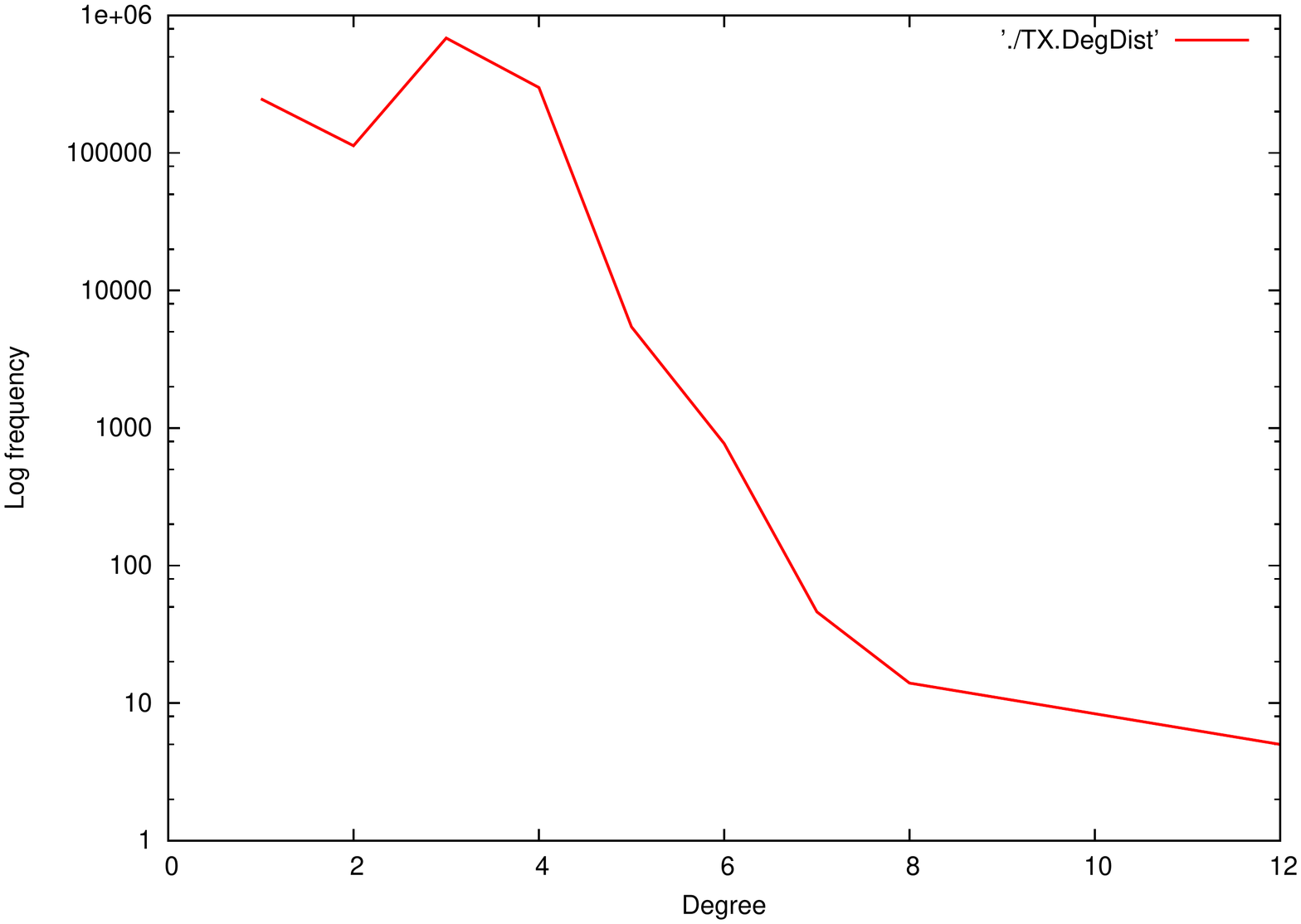}
\end{center}
\caption{Degree distribution for roadNet-TX network, in both log-log and linear-log
plots, showing a nearly Poisson degree distribution that falls off sharply 
with no power-law tail.}
\end{figure}

\begin{table}
\begin{center}
\subfigure[Social Networks]{
\label{fig:smSN1}
\begin{tabular}{|c|c|c|c|c|c|c|c|c|}
\hline
Network  & $|V|$ & $|E|$ & diam. & $\e d(x,y)$ & $\mean~c_v$ & med $c_v$ & $\delta_{\max}$ & $\e\delta(x,y)$ \\
\hline
\hline
\texttt{sn-Facebook} & 63392 & 816886 & 15 & 4.322 & 0.223 & 0.167 & 2.0 & 0.229 \\
\hline
\texttt{sn-Small} & 6116 	 &	  31374 	&	  15 	& 	  5.096&  0.243 	&  0.179& 3.0 & 0.304 \\
\hline
\texttt{sn-Medium} &   26567 	 &	226566 & 14 & 4.258 &  0.223 	&  0.167& 2.0 & 0.230 \\
\hline
\texttt{sn-Large} & 293501 &	  5589802 & 13 & 4.213 & 0.181 & 0.136 & 2.0 & 0.224\\
\hline

wiki-Vote & 7066 	 &  100736  & 7 & 3.248 & 0.142 	&  0.1 & 1.5 & 0.174 \\
\hline
\end{tabular}
}
\subfigure[Signed Networks]{
\label{fig:sm+-N1}
\begin{tabular}{|c|c|c|c|c|c|c|c|c|}
\hline
Network  & $|V|$ & $|E|$ & diam. & $\e d(x,y)$ & $\mean~c_v$ & med $c_v$& $\delta_{\max}$ & $\e\delta(x,y)$ \\
\hline
\hline
\texttt{soc-sign-Epinions} & 63192  & 648634	& 14	& 3.623 	& 0.312 & 0.167& 1.5 & 0.365	\\
\hline
\texttt{Slashdot081106}  & 49166  & 440370 & 11 & 3.584 &  0.0903  & 0.00735 & 1.5 & 0.386 \\ 
 \hline
\texttt{Slashdot090216}  & 53405  & 469210 & 12 & 3.646 &  0.0937  & 0.00647 & 1.5 & 0.386 \\
 \hline
\texttt{Slashdot090221}  & 53592  & 471933 & 12 & 3.645 &  0.0942  & 0.00654 & 1.5 & 0.385 \\ 
 \hline
\texttt{soc-Slashdot0811} & 77360 	 &	  546487 &  12  & 4.024 &  0.0555 	&  0.0& 1.5 & 0.191 \\
\hline
\texttt{soc-Slashdot0922} & 82168 	& 	  582533  &  13   & 4.069  &	  0.0603 	&  0.0& 1.5 & 0.191\\
\hline
\end{tabular}
}
\subfigure[RocketFuel Networks]{
\label{fig:smRF1}
\begin{tabular}{|c|c|c|c|c|c|c|c|c|}
\hline
Network  & $|V|$ & $|E|$ & diam. & $\e d(x,y)$ & $\mean~c_v$ & med $c_v$ & $\delta_{\max}$ & $\e\delta(x,y)$ \\
\hline
\hline
\texttt{1221 Telstra}   & 2998  & 3806 & 12 & 5.525 &  0.0141  &  0.0& 2.0 & 0.0987 \\ 
\hline
\texttt{1239  Sprintlink} & 8341  & 14025 & 13 & 5.184 &  0.0247  &  0.0& 2.0 & 0.236 \\ 
\hline
\texttt{1755  Ebone} & 605  & 1035 & 13 & 5.960 &  0.0510  &  0.0& 2.5 & 0.276 \\ 
\hline
\texttt{2914  Verio} & 7102  & 12291 & 13 & 6.0422 &  0.0691  &  0.0& 2.5 & 0.258 \\ 
\hline
\texttt{3257  Tiscali} & 855  & 1173 & 14 & 5.300 &  0.0133  &  0.0& 2.0 & 0.139 \\ 
\hline
\texttt{3356  Level3} & 3447  & 9390 & 11 & 5.0693 &  0.0854  &  0.0& 2.0 & 0.158 \\ 
\hline
\texttt{3967  Exodus} & 895  & 2070 & 13 & 5.944 &  0.177  &  0.0& 2.5 & 0.320 \\ 
\hline
\texttt{4755  VSNL} & 121  & 228 & 6 & 3.196 &  0.0372  &  0.0& 2.0 & 0.071 \\ 
\hline
\texttt{7018  AT\&T} & 10152  & 14319 & 12 & 6.947 &  0.00580  &  0.0& 2.5 & 0.251  \\ 
\hline
\end{tabular}
}
\subfigure[Peer-to-peer Networks]{
\label{fig:smP2P1}
\begin{tabular}{|c|c|c|c|c|c|c|c|c|}
\hline
Network  & $|V|$ & $|E|$ & diam. & $\e d(x,y)$ & $\mean~c_v$ & med $c_v$ & $\delta_{\max}$ & $\e\delta(x,y)$ \\
\hline
\hline
\texttt{p2p-Gnutella04}  & 10876  & 39994 & 10 & 4.636  &  0.00622  &  0.0& 2.5 & 0.288 \\ 
\hline
\texttt{p2p-Gnutella05}  & 8842  & 31837 & 9 & 4.596  &  0.00720  &  0.0& 2.0 & 0.292 \\ 
\hline
\texttt{p2p-Gnutella06}  & 8717  & 31525 & 10 & 4.572   &  0.00668  &  0.0 & 2.5 & 0.290 \\ 
\hline
\texttt{p2p-Gnutella08}  & 6299  & 20776 & 9 & 4.643  &  0.0109  &  0.0& 2.0 & 0.296 \\ 
\hline
\texttt{p2p-Gnutella09}  & 8104  & 26008 & 10 & 4.767  &  0.00954  &  0.0& 2.5 & 0.298 \\ 
\hline
\texttt{p2p-Gnutella24}  & 26498  & 65359 & 11 & 5.418  &  0.00551  &  0.0 & 2.5 & 0.294 \\ 
\hline
\texttt{p2p-Gnutella25}  & 22663  & 54693 & 11 & 5.545  &  0.00531  &  0.0 & 2.5 & 0.325 \\ 
\hline
\texttt{p2p-Gnutella30}  & 36646  & 88303 & 11 & 5.750  &  0.00630  &  0.0& 2.5 & 0.319 \\ 
\hline
\texttt{p2p-Gnutella31}  & 62561  & 147878 & 11 & 5.936  &  0.00547  &  0.0& 2.5 & 0.314 \\
\hline
\end{tabular}
}

\end{center}
\caption{Network Measures I}

\end{table}

\begin{table}
\begin{center}
\subfigure[Collaboration Networks]{
\label{fig:smCA1}
\begin{tabular}{|c|c|c|c|c|c|c|c|c|}
\hline
Network  & $|V|$ & $|E|$ & diam. & $\e d(x,y)$ & $\mean~c_v$ & med $c_v$& $\delta_{\max}$ & $\e\delta(x,y)$ \\
\hline
\hline
\texttt{CA-AstroPh}  & 17903  & 197031 & 14 & 4.194  &  0.633  &  0.667& 2.0 & 0.235 \\ 
\hline
\texttt{CA-CondMat}  & 21363  & 91342 & 15 & 5.352  &  0.642  &  0.700& 2.5 & 0.286 \\ 
\hline
\texttt{CA-GrQc}  & 4158  & 13428 & 17 & 6.049  &  0.557  &  0.533& 3.0 & 0.377 \\ 
\hline
\texttt{CA-HepPh}  & 11204  & 117649 & 13 & 4.673  &  0.622  &  0.673& 2.0 & 0.252 \\ 
\hline
\texttt{CA-HepTh}  & 8638  & 24827 & 18 & 5.945  &  0.482  &  0.333& 3.0 & 0.338 \\
\hline
\end{tabular}
}
\subfigure[Web Networks]{
\label{fig:smWeb1}
\begin{tabular}{|c|c|c|c|c|c|c|c|c|}
\hline
Network  & $|V|$ & $|E|$ & diam. & $\e^\star d(x,y)$ & $\mean~c_v$ & med $c_v$ & $\delta_{\max}$ & $\e\delta(x,y)$ \\
\hline
\hline
\texttt{web-Google} & 855802 & 4291352 & 24 & 6.334 & 0.605 & 0.519 & 2.0 &  0.234 \\
\hline
\texttt{web-BerkStan} & 654782 & 6581871 & 208 & 7.106 & 0.615 & 0.607 & 2.0 &  0.306 \\
\hline 
\texttt{web-Stanford} &  255265 & 1941926 & 164 & 6.815 & 0.619 & 0.667 & 1.5 &  0.198 \\
\hline
\end{tabular}
}

\subfigure[Road Networks.  ]{
\label{fig:smRoad1}
\begin{tabular}{|c|c|c|c|c|c|c|c|c|}
\hline
Network  & $|V|$ & $|E|$ & diam. & $\e^\star d(x,y)$ & $\mean~c_v$ & med $c_v$& $\delta_{\max}$ & $\e\delta(x,y)$ \\
\hline
\hline
\texttt{roadNet-CA} & 1957027 & 2760388 &  850 & 312.81 &  0.0464 & 0.0465 & 208.5 &  40.990\\
\hline
\texttt{roadNet-PA} & 1087562 & 1541514 & 782 & 310.51 &  0.0465 & 0.0384 & 195.5 &  35.013 \\
\hline 
\texttt{roadNet-TX}  & 1351137 & 1879201 &  1049 & 416.35 &  0.0470 & 0.0384 & 222.0 &  54.198 \\
\hline
\end{tabular}
}
\end{center}
\caption{Network Measures II}
\end{table}

\paragraph{}
To summarize, publicly available data from
communication and social networks 
describe networks ranging from a few hundred to a few millions vertices
and edges, with diameters ranging from 6 to 24, all well within the range of
the logarithm of the network size (as measured by the number of vertices). 
The degree distributions of these graphs, with two representative samples 
shown in Figures~7, also confirm the well-studied fact that some form of 
power-law distribution, at least in some range, provides a more accurate 
description of the nodal degrees of many graphs than distributions with 
exponential tails.  The clustering coefficients measured are, however,
not always large as seen in Tables~1-2.  Collectively, these observations 
are consistent with past findings about the small world and degree
distribution of large-scale networks. 
 
Simultaneously, Tables~1 and 2 show that the maximal (4-point) $\delta$ that 
we measured never exceeded $3$ and in the majority of cases $2.5$, with the 
average sampled $\delta$ considerably smaller.  
Based on the discussions in Sections~2.2-2.3, these findings provide 
comfortable ground for presumption of hyperbolicity for the communication
and social networks.  The road networks studied provide a counter-balance
to the above observations, but their 
characteristic parameters are completely consistent with a flat, zero
curvature (or $\delta=\infty$) geometry as indicated by similarity of their
curvature plots, Figure~\ref{fig:curvRN}, and that of the square grid,
Figure~\ref{subfig3-2}.  
We conclude that hyperbolicity is a general feature of large-scale networks 
that is distinct from specific features of degree distributions.
An updated version of the taxonomy chart in~\cite{NS11}, showing the 
features of various networks and the relationship between different features
is in Figure 9.

\begin{figure}
\begin{center}
\includegraphics[angle= 0, width =3.0in]{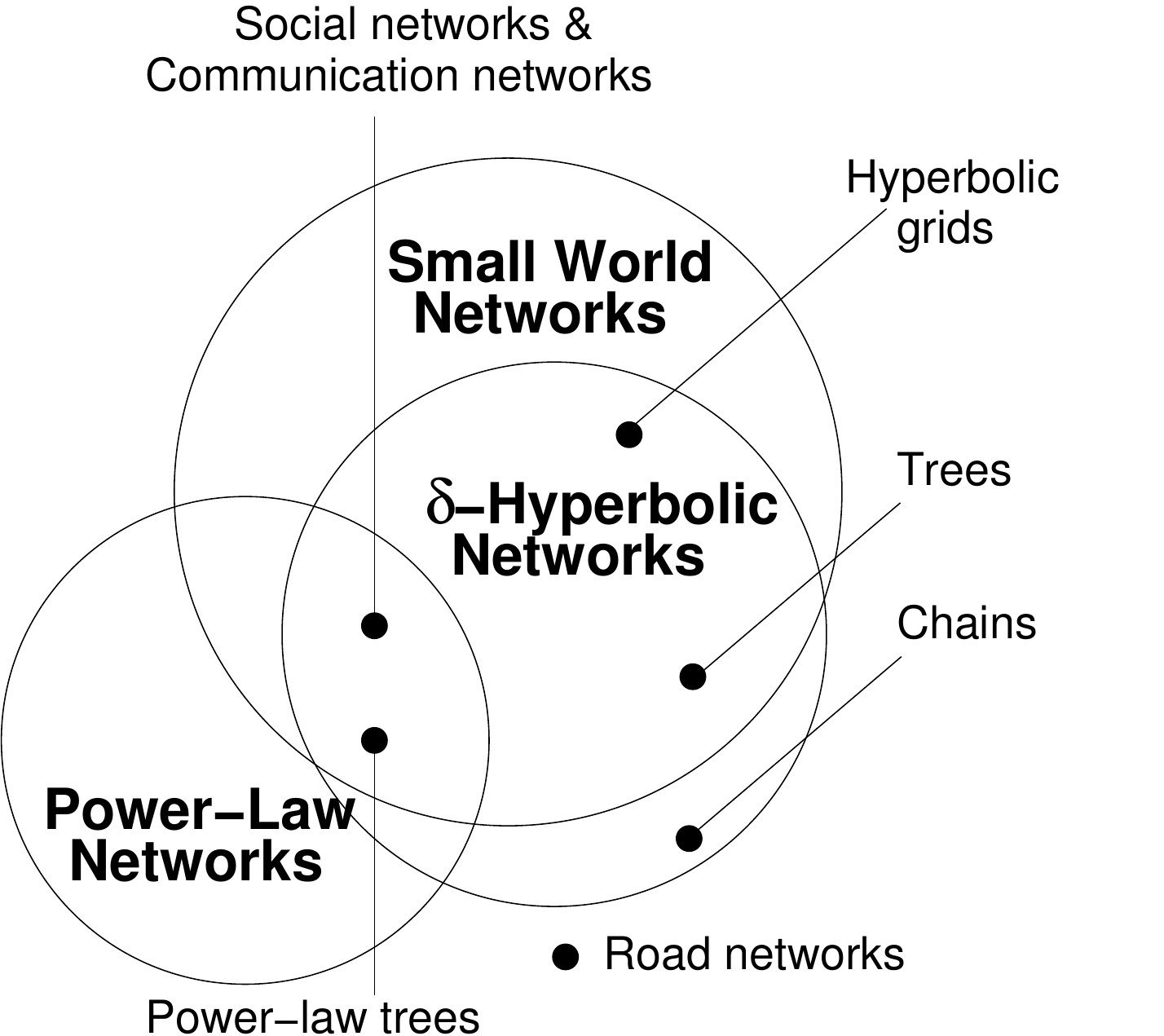}
\caption{Extension of the taxonomy of networks originally 
proposed in~\cite{NS11}, based on measurements shown in Tables~9-10.}
\label{taxonomy}
\end{center}
\end{figure}

\section{Interpreting Curvature Plots -- Evidence for Hyperbolicity}
\label{sec:evidence}

\begin{figure}[htb]
\begin{center}
\includegraphics[width = 5in]{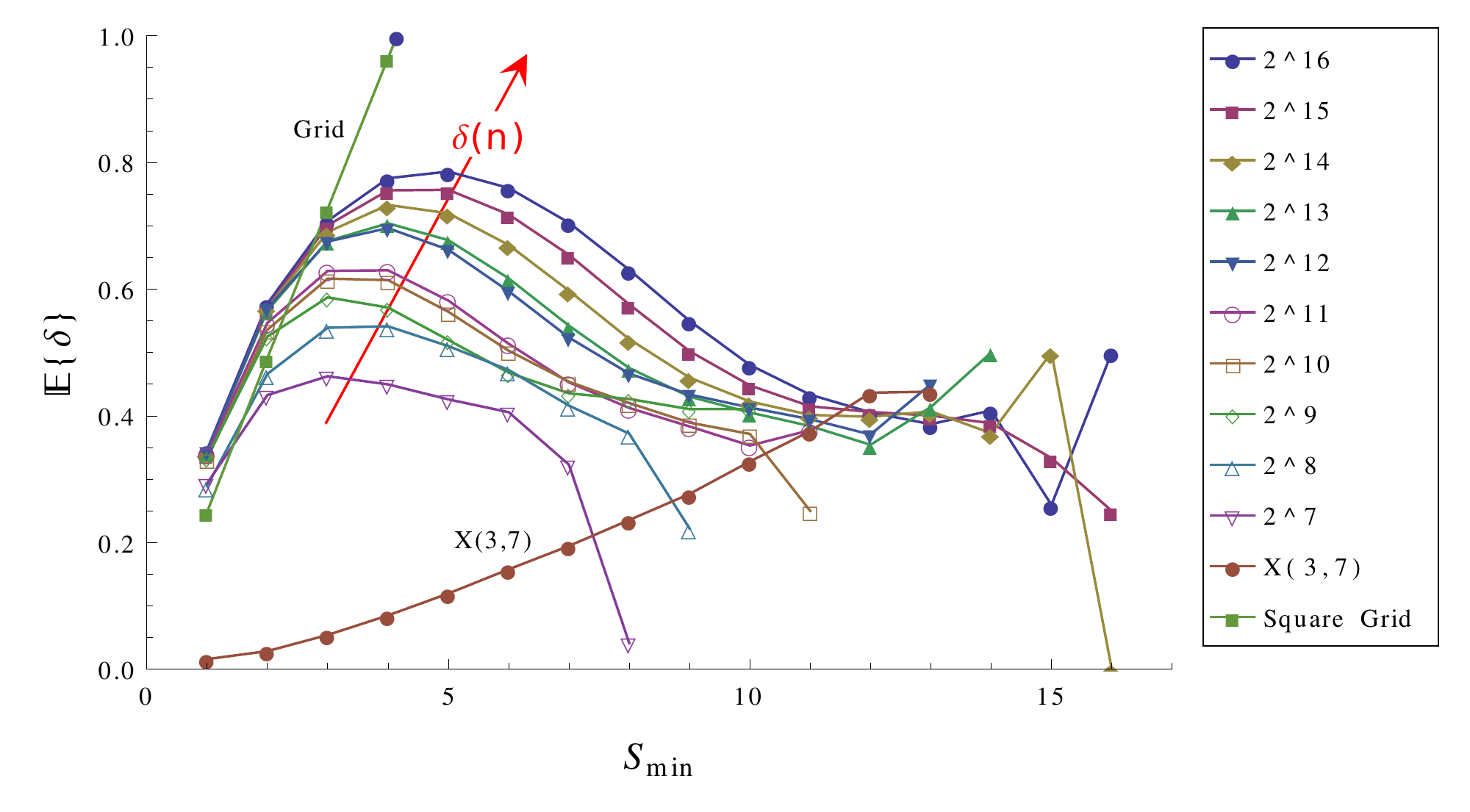}
\caption{Curvature plots of random graphs $G_{2^k,p}$, $k = 7,8,...,16$ and $p=3/n$, compared to standard reference curvature plots of $\mathbb{H}_{3,7}$ (labelled $X(3,7)$) and the Square Grid.
}
\label{fig:e-rCPs}
\end{center}
\end{figure}

In Figure~\ref{fig:H37} we see a good illustration of the growth of $E[\delta]$
as a function of quadrilateral (or triangle) size, for a sequence of 
truncated \emph{bona fide} 
hyperbolic grids $\mathbb{H}_{3,7}$.  Here we see a regime of clear growth followed by
abrupt flattening, regardless of the size of the base quadrilateral (triangle),
the maximum of which reflects the size of the graph, $n$.  
By contrast, Figure~\ref{fig:squareGrid} shows $E[\delta]$ increases strictly
monotonically as a function of the size of a square grid for which triangles
clearly are not thin.

In interpreting flattening in a curvature plot (CP), 
one has to be careful about boundary
effects. Figure~\ref{fig:e-rCPs} shows multiple CPs for different sizes 
of the Erd\"os-Renyi
random graph. For each curve, we observe a rise in the value of $E[\delta]$ followed by 
a decline. 
However, there is a net increase in $\max E[\delta]$ as the network $n$ size gets 
larger.
This ``rise to the north-east'' of the CP peak (or hump) is a tell-tale
sign that the decline or flattening of each curvature plot is not indicative of hyperbolicity, but
is in fact a boundary effect.  
In~\cite{NST12},
it is shown rigorously that the Erd\"os-Renyi random graphs in the regime studied here
are not $\delta$-hyperbolic. By contrast, the ``rise followed by flattening" seen in the curvature plots of 
Figure~\ref{fig:H37} is not a boundary effect since the height of the maximal $E[\delta]$
for any size graph remains constant, and constitutes strong evidence for hyperbolicity. 

Thus a flattening or declining $E[\delta]$ may be interpreted in one of two ways:
either an indication that the graph is hyperbolic, or as a boundary effect. 
But with careful considerations, one is able to use curvature plots
to see if $E[\delta]$ is either constant or growing across a
family of graphs and distinguish between these two. As in all numerical detection algorithms, detection of
the ``rise'' in the curvature plot has inevitable limits. If $\max E[\delta]\sim \log n,$
one needs much larger instances of graphs before the ``north-east" trend is detectable, as
is seen in Figure~\ref{fig:e-rCPs}. If $\max E[\delta]$ grows even more slowly with $n,$
the situation is still worse; if 
(say) $\max E[\delta] \sim \log \log n,$
detection will be effectively impossible with any numerical scheme for
any reasonable network size (even $\log \log (10^{100}) =2$).
However, in such a case, even though $\delta(n)$ is mathematically unbounded, we may
reasonably characterize such a network as being 
``practically hyperbolic''.

\begin{figure}
\begin{center}
\subfigure[Facebook Social Networks]{
\label{fig:curvFB}
\includegraphics[width=7.5cm]{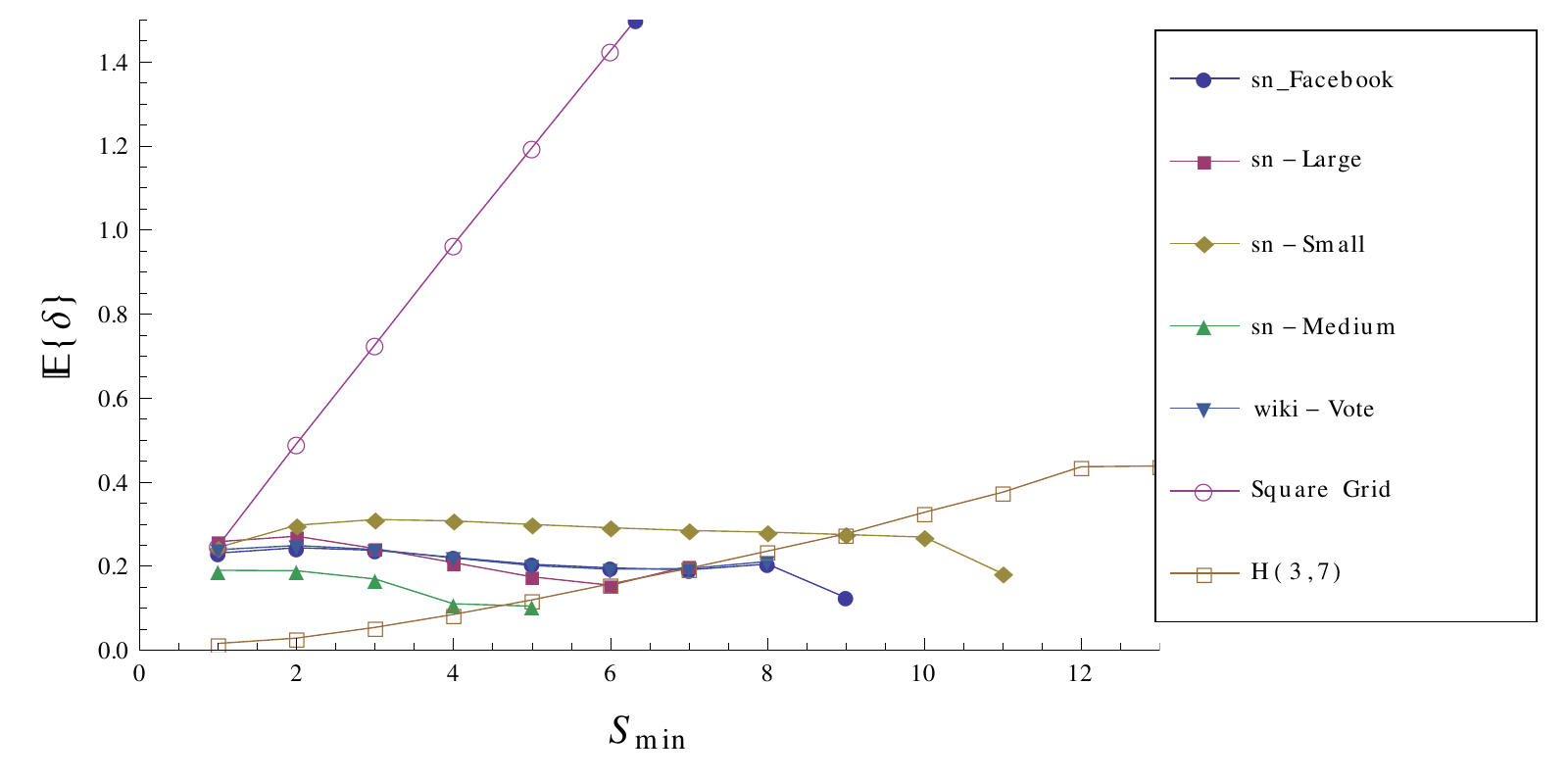}
\label{fig:NCP}
}
\subfigure[Signed Networks]{
\label{fig:curvSN}
\includegraphics[width=7.5cm]{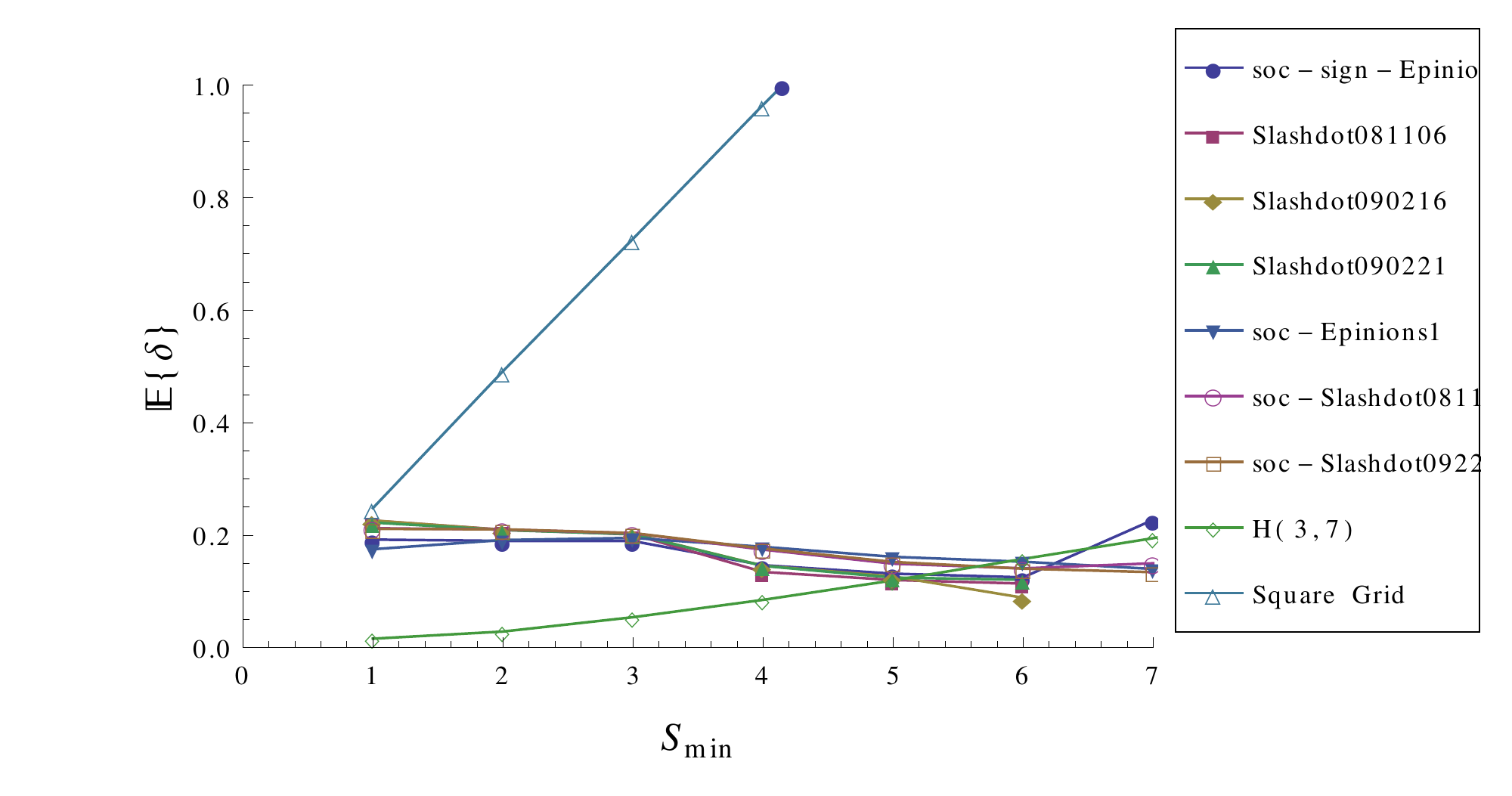}
}
\subfigure[RocketFuel Networks]{
\label{fig:curvRF}
\includegraphics[width=7.5cm]{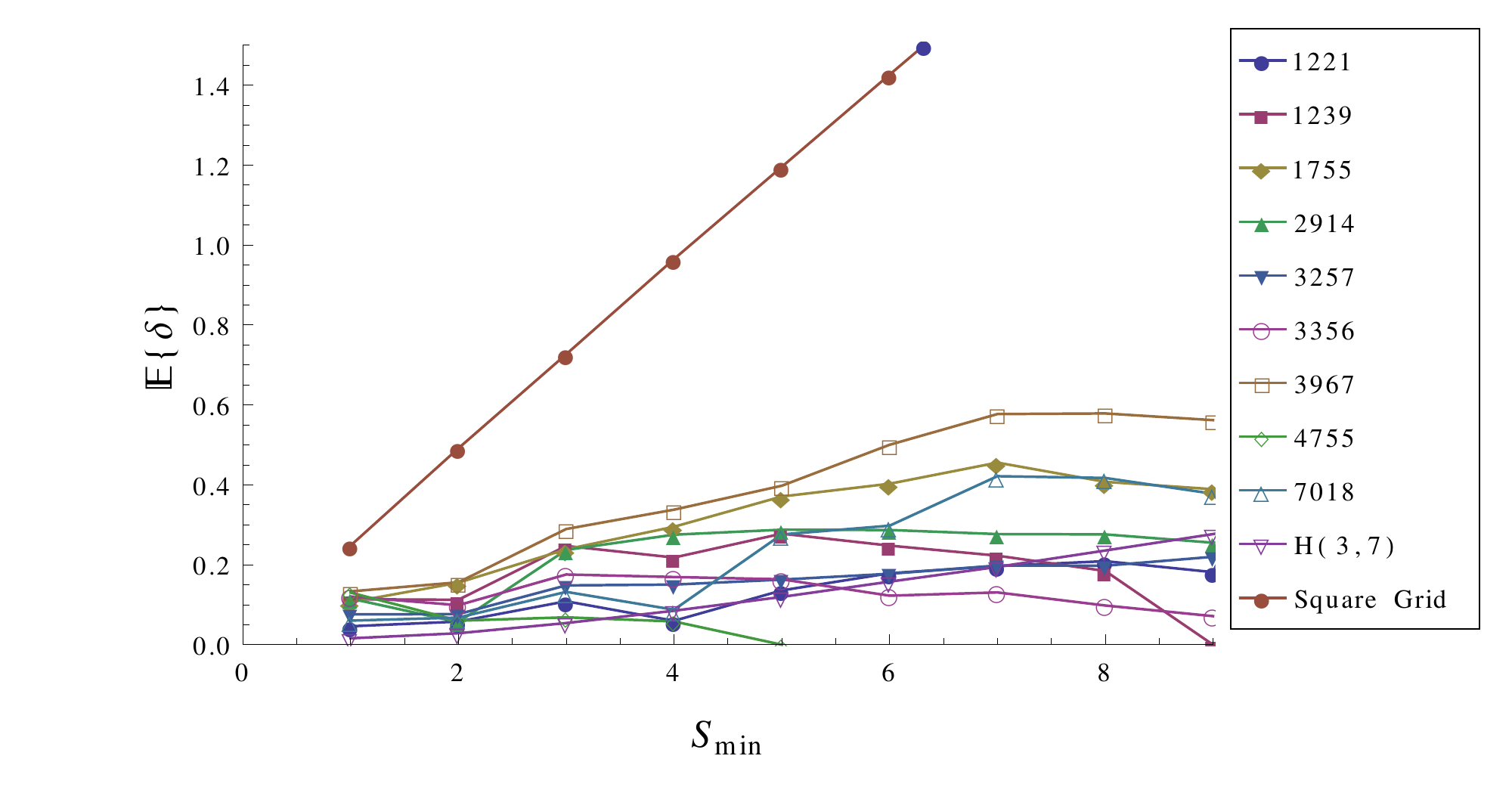}
}
\subfigure[Peer-to-peer Networks]{
\label{fig:curvP2P}
\includegraphics[width=7.5cm]{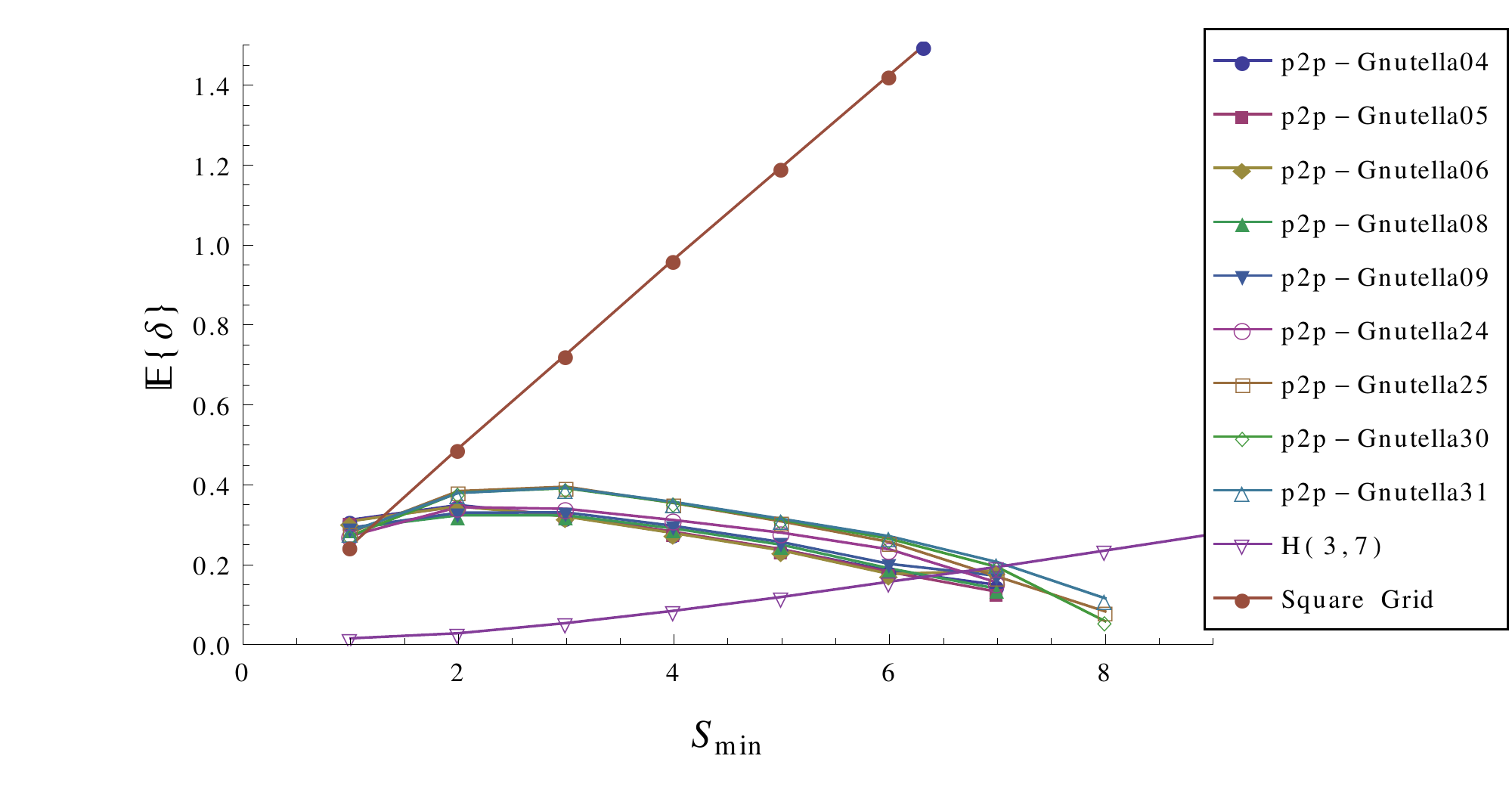}
}
\subfigure[Collaboration Networks]{
\label{fig:curvCN}
\includegraphics[width=7.5cm]{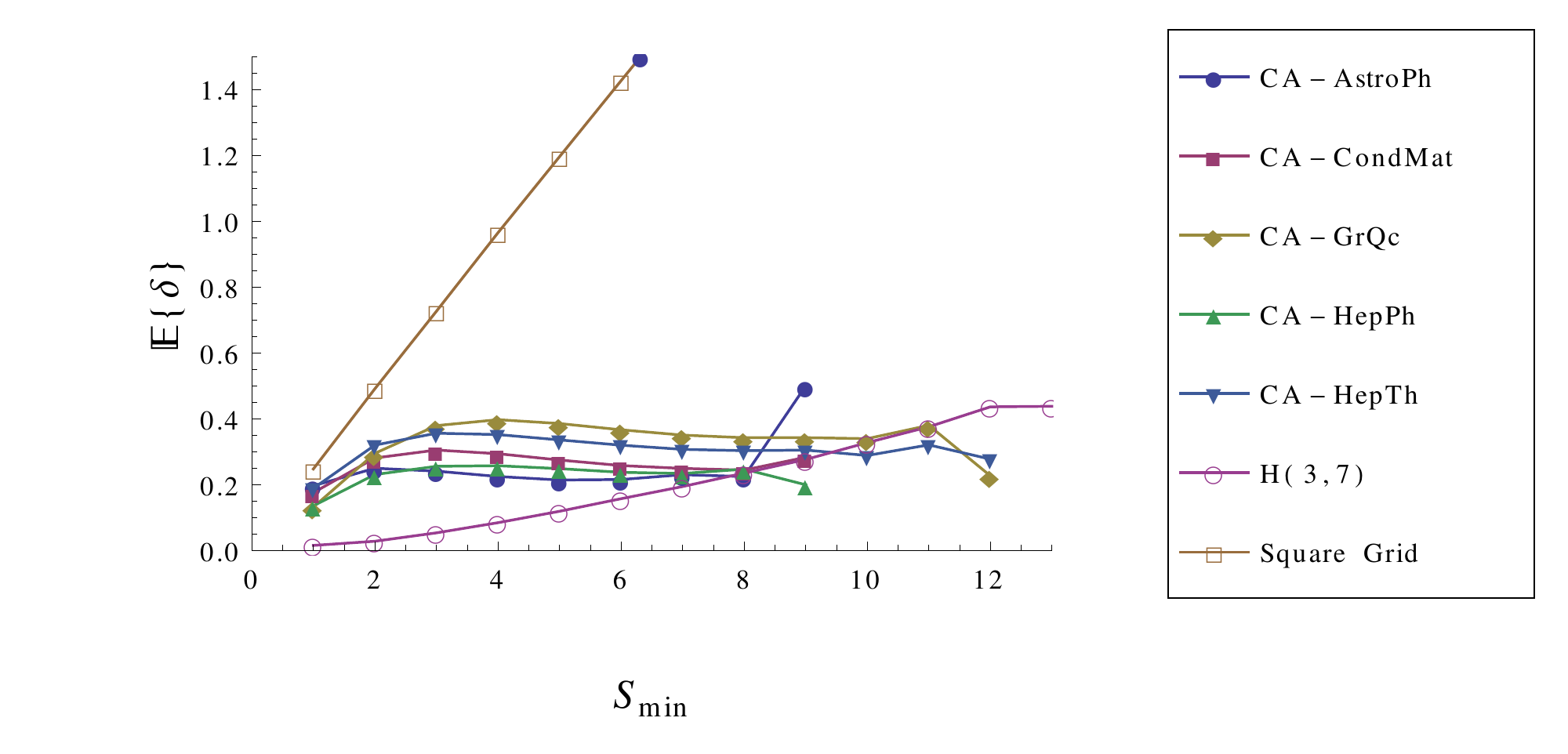}
}
\subfigure[Web Networks]{
\label{fig:curvWN}
\includegraphics[width=7.5cm]{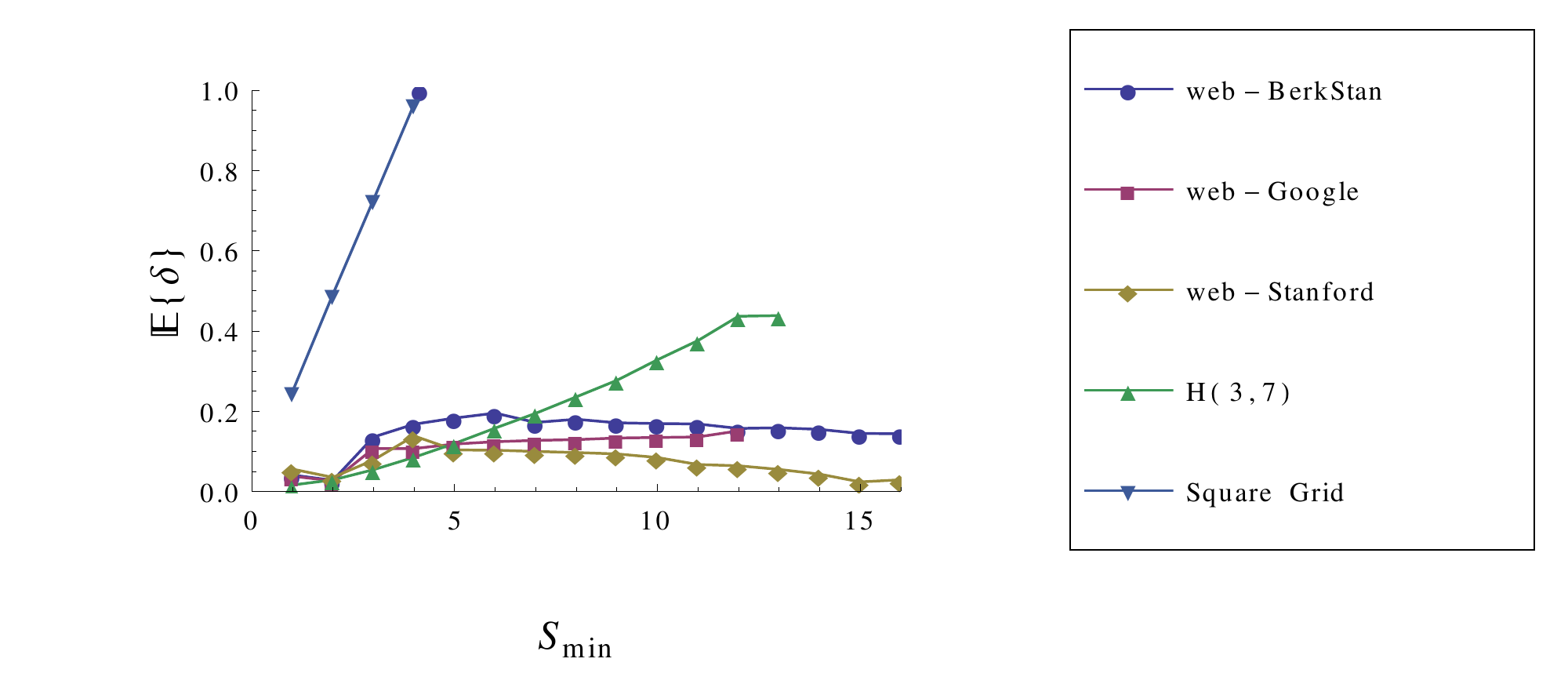}
}
\subfigure[RoadNetworks]{
\label{fig:curvRN}
\includegraphics[scale=0.75]{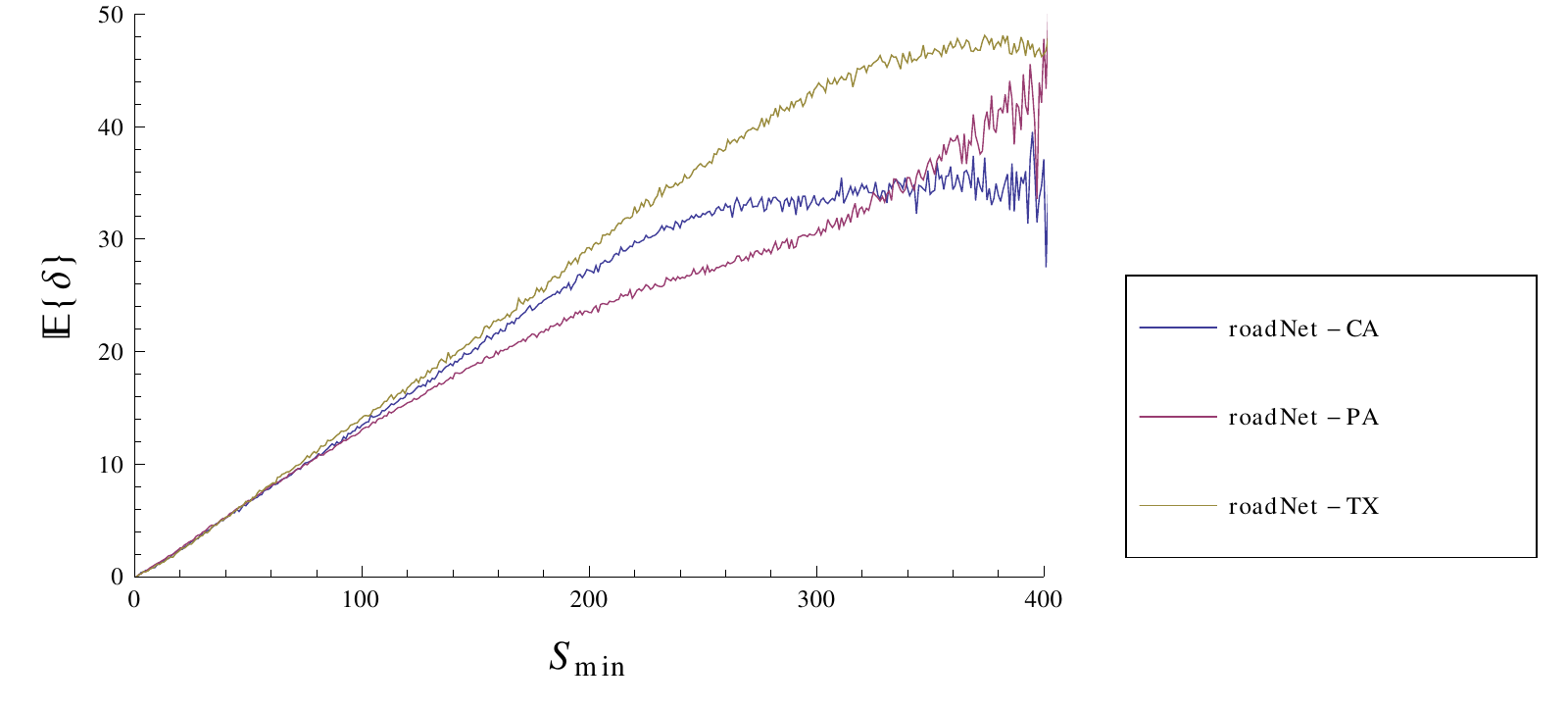}
}
\caption{Network Curvature Plots}
\end{center}
\end{figure}

For real networks, which typically do not come from a family of graphs
with specific and common characteristics, as the above examples did, reading and
interpreting a curvature plot may involve additional subtleties, which we
described in the context of the dozens of real networks we studied.  
To eliminate the inevitable discrepancy in scaling of the curvature 
plot axes
for different classes and sizes of networks, we used the
hyperbolic and Euclidean grid curvature plots to bound and calibrate the (relative) 
size and growth of $E[\delta]$ for each curvature plot, 
as shown for example in Figure~\ref{fig:e-rCPs}.
Thus we see the road network Figure~\ref{fig:curvRN} is clearly non-hyperbolic as 
$E[\delta]$ grows similarly (but with
different slope) to the square grid.  For all other networks, 
we either see no growth 
in $E[\delta]$, as in Facebook Figure~\ref{fig:curvFB} and 
SignedNetwork Figure~\ref{fig:curvSN}, 
or a gentle rise followed by
a leveling off or even a decrease in the mean value of $E[\delta]$, 
as in Peer-to-PeerFigure Figure~\ref{fig:curvP2P}, 
Collaboration Figure~\ref{fig:curvCN}, 
Web Networks Figure Figure~\ref{fig:curvWN} and Rocketfuel Figure~\ref{fig:curvRF}. 
The flattening in the right tail of the plots in Figure~\ref{fig:curvRN} 
is due to the boundary effects discussed above, while the jaggedness is due to a lack 
of sufficient statistics
for larger quadrilaterals. 

\section{Scaling of Curvature Plots -- Renormalization of Networks}
\label{sec:rg}
In the previous sections, we have seen how to measure the curvature
of a network, and provided evidence that many well-known networks
are hyperbolic. For extremely large networks (in the order of 
billions of nodes), 
measuring the curvature
is computationally expensive. We note that the problem of hyperbolicity
detection has not been studied much before, but see~\cite{INRIA} for a 
recent discussion of algorithms for traditional approaches to this 
problem for graphs with up to 10K nodes.  In this section, we discuss how to
reduce the size of a network graph by thinning the nodes, that is, by 
merging adjacent nodes until the network size reduces to the 
level when computing the curvature is tractable. We do this via
the process known as the renormalization group.  In order for 
this kind of thinning to be meaningful, the curvature has to 
be preserved by the thinning process. 
We provide arguments and results for networks whose curvature is known
to support this assertion. 

The renormalization group is a semi-group of transformations that coarse
grains any large interacting system by eliminating (`integrating out')
features at small length scales and simultaneously changing the unit
of length. In the function-space of all probability distributions,
limiting distributions of probability theory are fixed points of the
corresponding renormalization group.  To put the approach into a clear 
context, we start with a traditional example of spin chains.
The simplest example is that of a long chain of spins,
$S_i = \pm 1$
for $i = 1, 2, \ldots 2^n.$ The probability of any spin configuration is of the
form
\begin{equation}
p(S_1,S_2,\ldots S_{2^n}) \propto \exp[J\sum S_i S_{i+1}].
\label{renorm1}
\end{equation}
If one calculates the reduced probability distribution for all the
even spins, this is equivalent to integrating out all the odd spins.
It is easy to show that the result is a new spin chain with a
probability distribution that is similar in form to Eq.(\ref{renorm1})
but with $\exp[2 J^\prime] = \cosh 2 J.$ Spins which were a distance
$x$ apart are now a distance $x/2$ apart.  Iterating this process,
after the $m$'th iteration we have a chain of length $2^{n-m}.$ The
parameter $J$ decreases at each iteration, approaching zero as the
number of iterations becomes infinite. Accordingly, the joint
probability distribution for two spins that are far apart is the
same as for adjacent spins with very small $J,$ i.e. they are
essentially uncorrelated.

One of the attractive features of the renormalization group is its
property of {\it universality}. A large family of physical systems
(for our purpose here, probability distributions in phase space)
converge to the same universal behavior under renormalization. Thus
the properties of a system at large length scales do not depend on
inessential features of its small scale structure.  Since we are
interested in the properties common to large communication networks,
this is very useful.

There have been various studies of network renormalization
earlier~\cite{Makse,Paczuski,N03}.  If the network is embedded
in a simple manifold~\cite{N03}, it is straightforward to lay
down boxes and merge all the nodes inside a box, in a standard real
space renormalization group prescription. If there is no such
embedding, one can still partition the network into disjoint clusters
such that any pair of nodes inside a cluster are no more than a
distance $l$ apart~\cite{Makse}, but the partitioning is not unique.
Alternatively, one can pick a seed node at random and fuse a cluster
of nodes centered at the seed and separated by no more than a
distance $l,$ then repeat this process with successive seed
nodes~\cite{Makse,Paczuski}. Various properties of the network such 
as its small world behavior (or lack thereof) and its degree distribution
are shown to scale in the same manner as in the standard scaling theory
of phase transitions.

We apply this approach to the study of network curvature. We choose
a node and merge it with one of its neighbors. Multiple edges between
pairs of nodes, and the edge connecting the new merged node to
itself, are eliminated.  Rescaling is achieved by assuming that all
new edges are still of unit length. This process is repeated. If
nodes and the neighbors with which they are merged are chosen
randomly each time, as the renormalization proceeds there is
substantial variation in the number of nodes in the original network
that are fused into each node of the renormalized network. To reduce
this effect, at each stage we keep track of the number of original
nodes in each renormalized node, and pick a renormalized node
randomly from the set for which this is smallest, and merge it with
its neighbor for which this is smallest. We construct curvature
plots --- as described in the previous sections --- at various
stages in the renormalization process, when the number of nodes in
the network is reduced from $N$ to $N/2^n$ with different values
of $n.$

\begin{figure}[htb]
\begin{center}
\includegraphics[width=3in]{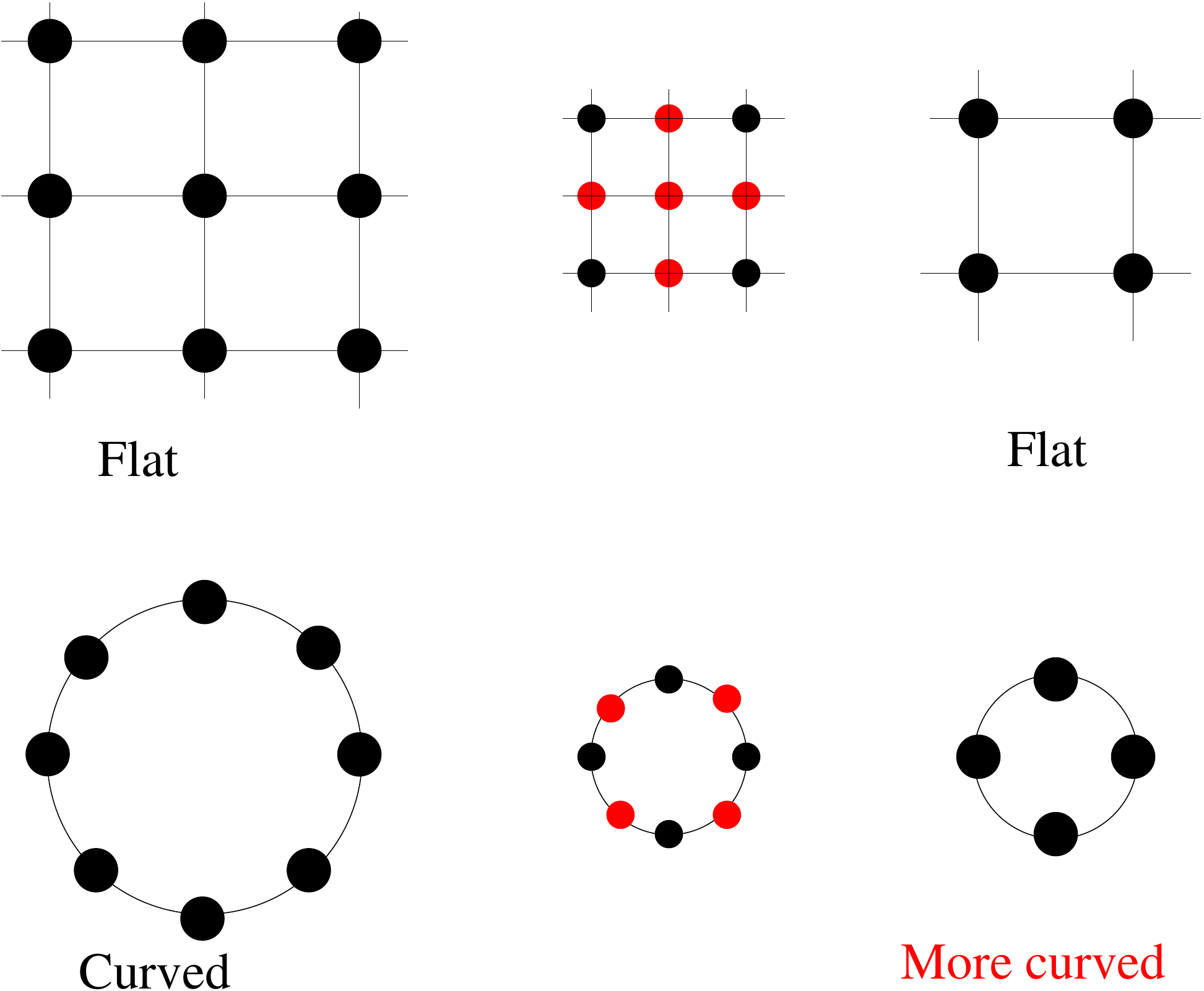}
\caption{Renormalization of a flat surface and the surface of a
sphere (shown in cross section), by eliminating nodes and rescaling.}
\label{fig:rgcurv}
\end{center}
\end{figure}
Before presenting the results, we first discuss what we might expect.
We consider the case
of a continuum manifold (with some lattice structure at the small
scale), with constant Gaussian curvature. Figure~\ref{fig:rgcurv}
shows examples of a flat surface and the surface of a sphere. After
rescaling and eliminating nodes, the flat surface clearly remains
flat, with a smaller size but the same lattice spacing. On the other
hand, the curvature of the spherical surface increases when it
becomes smaller. This suggests that differences in curvature of
different graphs are retained and in fact enhanced by renormalization.
This argument is not a proof: as we have shown~\cite{NS11}, Gaussian
curvature for a graph can be very different from $\delta$-hyperbolicity
as defined by Gromov; it is possible to alter a graph in a manner
that changes its discretized Gaussian curvature dramatically without
changing its $\delta$ parameter. In fact, the example above is that
of a spherically curved surface, whereas with $\delta$-hyperbolicity
we only distinguish whether a graph is negatively curved or not.
Accordingly, we have to rely on numerics to validate our intuition from
continuum manifolds.

\begin{figure}[htb]
\begin{center}
\includegraphics[width=3.1in]{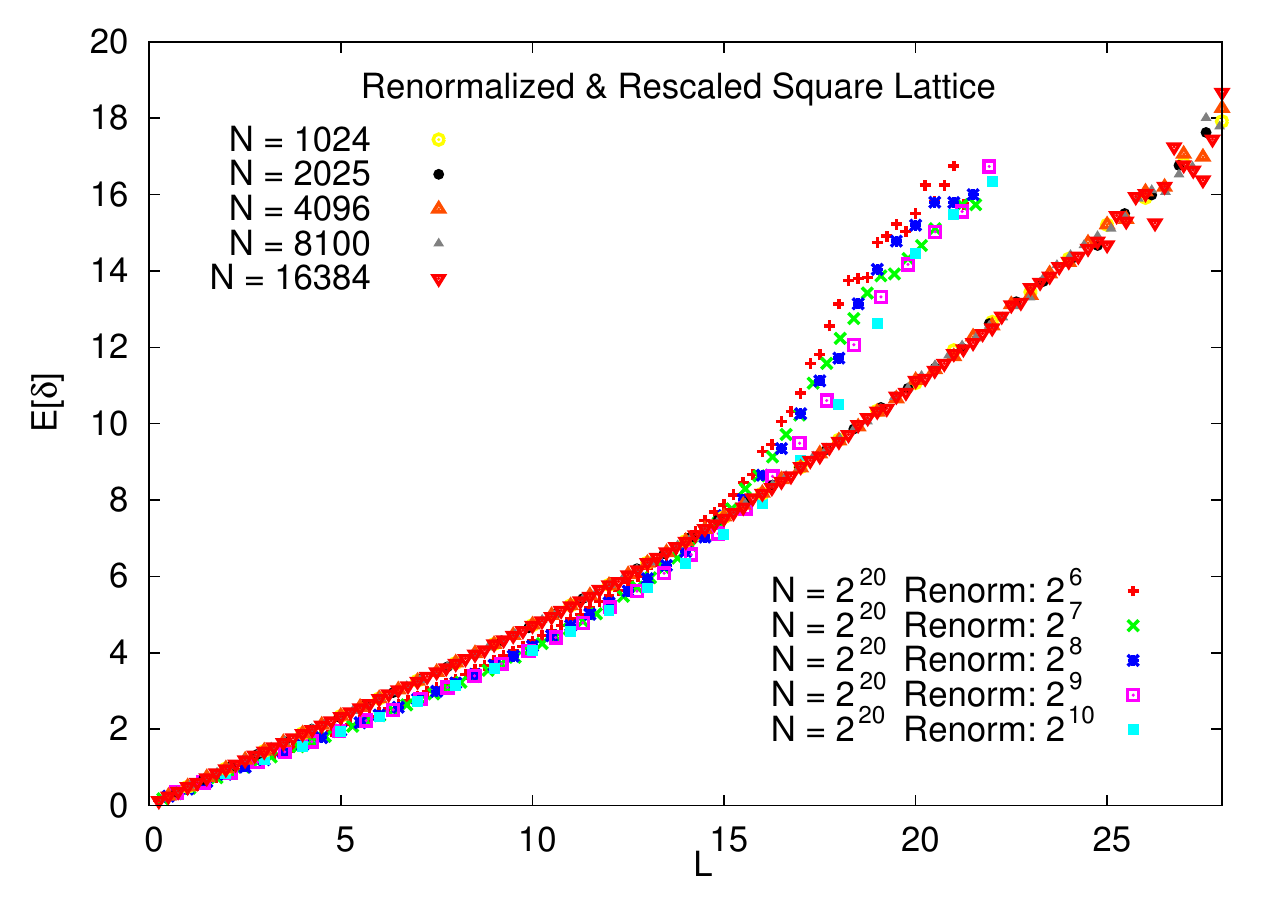}
\includegraphics[width=3.1in]{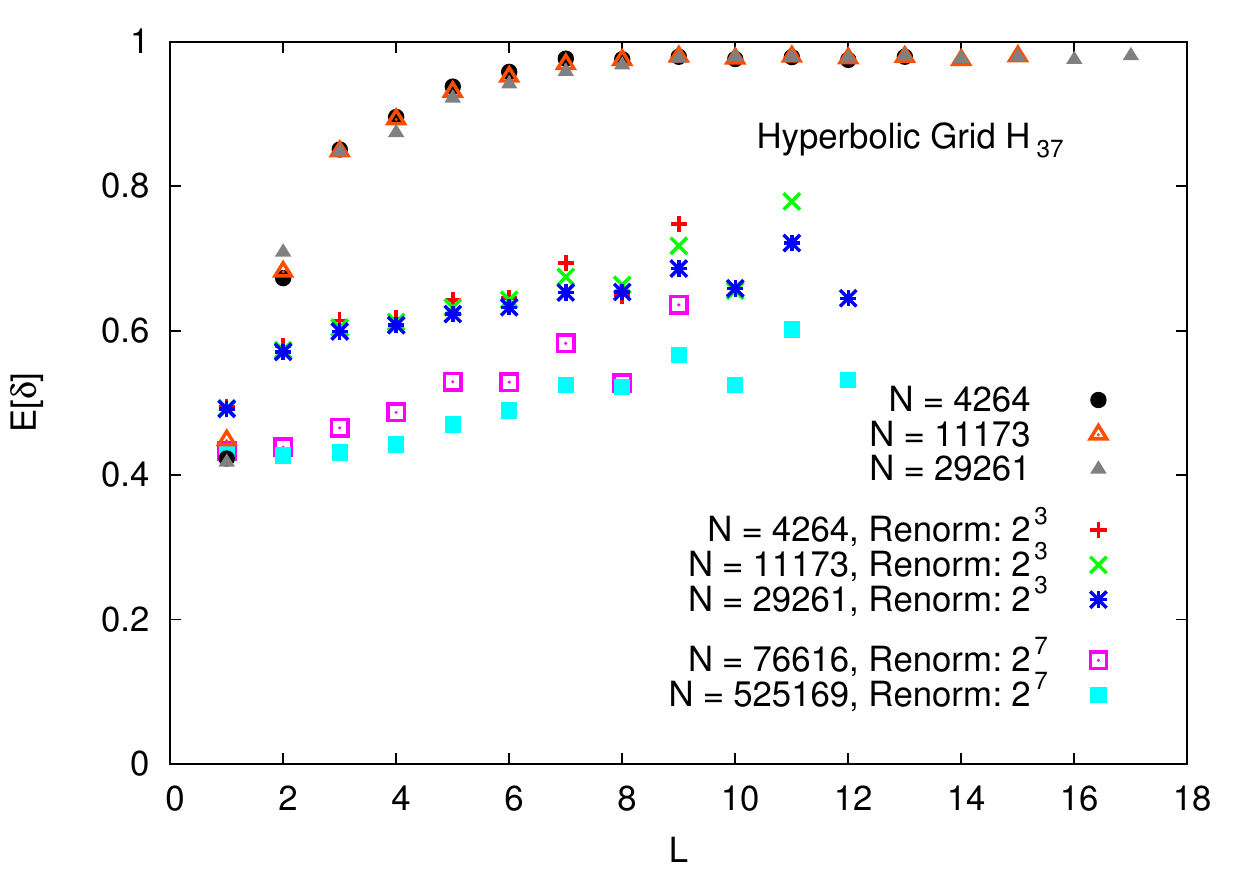}
\caption{(Left) Rescaled curvature plot for square lattices of different
sizes, and for renormalized versions of a square lattice with
$2^{20}$ nodes. (Right) Curvature plots for the $\mathbb{H}_{3,7}$ hyperbolic grid 
truncated to
various sizes, and for renormalized versions thereof.}
\label{fig:sqlat}
\end{center}
\end{figure}
The left panel of Figure~\ref{fig:sqlat} shows the results of this process for the
square lattice. The curvature plots for unrenormalized square
lattices with approximately $2^n$ nodes with $10\leq n \leq 14$ are
shown. To eliminate finite size effects, both axes in the figure
are multiplied by a factor of $2^{5 - n/2}$ to compensate for the
increase in the linear dimension $2^{n/2}$ of the graph as $n$ is
increased. For a graph that is not $\delta$-hyperbolic, this kind
of adjustment of the axes is expected~\cite{NS11} to collapse all
the curvature plots onto a universal curve, as is seen here.  The
figure also shows the curvature plots for a lattice with $N = 2^{20}$
nodes, renormalized to varying levels: the number of nodes in the
graph has been halved $6\leq m \leq 10$ times, and both axes have
been multiplied by $2^{m/2 - 5}.$ The curves are fairly close to
each other, but the collapse is of sufficient quality only in the
linear region. At the same time, it is clear that the curvature
plots of the renormalized graphs show that they are still not
$\delta$-hyperbolic.
By contrast, the right panel of Figure~\ref{fig:sqlat} shows curvature plots for the
$\mathbb{H}_(3,7)$ hyperbolic grid. All the nodes that are at a distance $\leq
r$ from a central node are included, with $r=7,8$ or 9, corresponding
to $N = 4264, 11173$ and 29261. The curvature plots for the
unrenormalized graphs lie on top of each other without any adjustments
to the axes, demonstrating that the flattening of the curvature
plots is due to $\delta$-hyperbolicity. After renormalization, the
curvature plots for all three graphs are still approximately
coincident, but with random scatter. However, it is clear that the
curvature plots of the renormalized graphs correspond to graphs
that are not $\delta$-hyperbolic, and that the flattening of the
curvature plots is enhanced by renormalization.

\begin{figure}[htb]
\begin{center}
\includegraphics[width=3.1in]{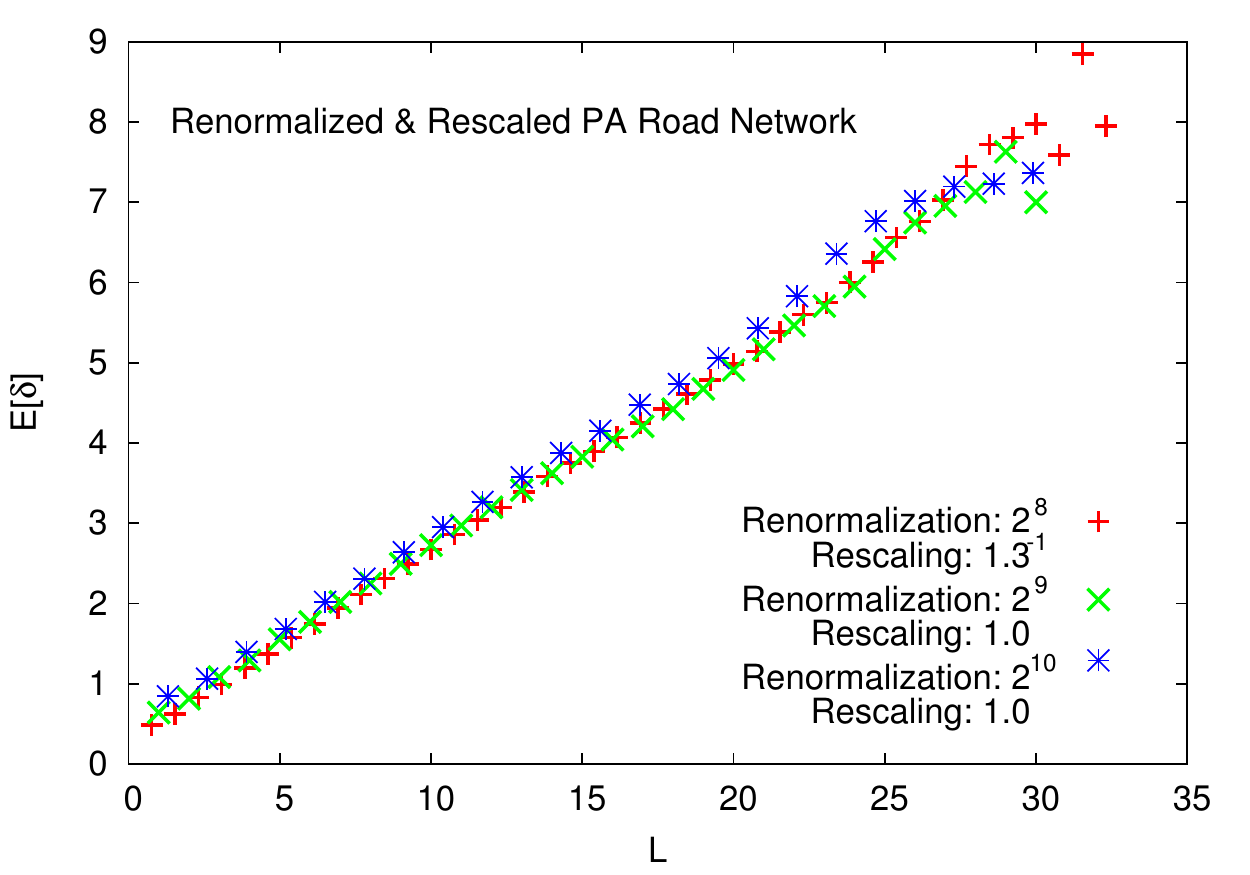}
\caption{Rescaled curvature plots for renormalized versions of the
Pennsylvania road network. The rescaling parameter was adjusted visually.}
\label{fig:PA_roads}
\end{center}
\end{figure}

We see that, consistent with our intuition from continuum manifolds,
$\delta$-hyperbolicity is preserved and enhanced when a graph is
renormalized, while a graph that is not $\delta$-hyperbolic remains
so.  Turning to physical networks, Figure~\ref{fig:PA_roads} shows
renormalized curvature plots for the Pennsylvania road network; the
original graph has $N > 10^6,$ and the curvature plot can be found
with difficulty and shows that the network is not $\delta$-hyperbolic.
After renormalizing down to $N/2^m$ nodes with $m=8, 9$ or 10, and
multiplying the axes by $\lambda^{m-9}$ (with the parameter $\lambda$
adjusted to 1.3), the curvature plots lie on top of each other.
This is similar to what was seen for the square lattice. Thus the
curvature plots for the (much smaller) renormalized graphs are
sufficient to show that the original graph is not $\delta$-hyperbolic.

\section{Concluding Remarks}
\label{sec:conclude}
We have used Curvature Plots~\cite{NS11} via the 3-point and 4-point conditions,
as defined by Rips and Gromov~\cite{Gromov87} to explore a novel large scale 
feature of a range of networks, including road, communication, citation,
collaboration, peer-to-peer, friendship and other social networks. Our computations
provide evidence that these graphs exhibit strong hyperbolicity.
For the 3-point test, this is in the sense that the average and maximum 
``thickness'' or $\delta$ 
of arbitrary triangles is significantly smaller than the graph diameters,
a characteristic that make them behave, in many respects, like trees 
(where the thickness or $\delta$ of a tree is simply zero).  The 4-point
test also shows the same thing by comparing distances for quadruples 
of points with corresponding distances for a tree.
We further showed that road networks, being typically nearly planar like two 
dimensional grids, are
naturally not hyperbolic. 

We also provided a renormalization mechanism
to facilitate and scale the detection of hyperbolicity in very large graphs;
using models of graphs for which it is known whether they are
hyperbolic, we demonstrated that this property is preserved by renormalization.

We have shown that hyperbolicity can coexist with other well-known
characteristics of large scale networks, namely a power-law degree distribution
and high clustering coefficient, as these are local characteristics which typically 
do not influence large-scale features of networks.  The large-scale 
feature of hyperbolicity,
in conjunction with such local characteristics,
provides a fuller picture of social networks.  
We conclude that the negative curvature of social networks is an additional 
and so far overlooked property which helps further classification of 
what is otherwise a bewilderingly complex array of natural 
and man-made networks, thus clarifying their underlying intrinsic
features.

\section{Acknowledgements}
The authors wish to thank Vladimir Marbukh for helpful discussions.
This work was supported by grants FA9550-11-1-0278 and 60NANB10D128 from AFOSR 
and NIST, respectively. The work of Kennedy was additionally supported by a 
postdoctoral grant from the Canadian NSERC. 

\bibliography{biblio}

\begin{thebibliography}{10}

\bibitem{SNAP}
Data archive at http://http://snap.stanford.edu/.

\bibitem{Rocketfuel}
Data archive at http://www.cs.washington.edu/research/networking/rocketfuel/.

\bibitem{BAJ00}
Albert-Laszlo Barabasi, Reka Albert, and Hawoong Jeong.
\newblock Scale-free charactetristics of random networks: the topology of the
  {W}orld-{W}ide {W}eb.
\newblock {\em Physica A: Statistical Mechanics and its Applications},
  281:69--77, 2000.

\bibitem{Paczuski}
G.~Bizhani, V.~Sood, M.~Paczuski, and P.~Grassberger.
\newblock Random sequential renormalization of networks: application to
  critical trees.
\newblock {\em Physical Review E}, 83(3):036110, 2011.

\bibitem{BCK09}
Marian Boguna, Dmitri Krioukov, and K.~C. Claffy.
\newblock Navigability of complex networks.
\newblock {\em Nature Physics}, 5(1):74--80, 01 2009.

\bibitem{BOW90}
B.H. Bowditch.
\newblock Notes on {G}romov's hyperbolicity criterion for path-metric spaces,
  1990.

\bibitem{BH99}
M.R. Bridson and A.~H{\"a}fliger.
\newblock {\em Metric Spaces of Non-Positive Curvature}.
\newblock Grundlehren der mathematischen Wissenschaften. Springer, 1999.

\bibitem{Buneman74}
Peter Buneman.
\newblock A note on the metric properties of trees.
\newblock {\em Journal of Combinatorial Theory, Series B}, 17(1):48 -- 50,
  1974.

\bibitem{INRIA}
Nathann Cohen, David Coudert, and Aur\'{e}lien Lancin.
\newblock Exact and approximate algorithms for computing the hyperbolicity of
  large-scale graphs.
\newblock {\em Publications of INRIA, hal-00735481--version 4,
  Sophia-Antiplos}, March, 2013.

\bibitem{dlH00}
P.~de~la Harpe.
\newblock {\em Topics in Geometric Group Theory}.
\newblock Chicago Lectures in Mathematics. University of Chicago Press, 2000.

\bibitem{FFF09}
Mihallis Faloutsos, Petros Faloutsos, and Christos Faloutsos.
\newblock On power-law relationships of the internet topology.
\newblock {\em ACM SIGCOMM}, 29(4), 1999.

\bibitem{GKBM10}
M.~Gjoka, M.~Kurant, C.T. Butts, and A.~Markopoulou.
\newblock Walking in {F}acebook: A case study of unbiased sampling of
  {O}{S}{N}s.
\newblock In {\em INFOCOM, 2010 Proceedings IEEE}, pages 1--9, 2010.

\bibitem{Gromov87}
M.~Gromov.
\newblock Hyperbolic groups.
\newblock In {\em Essays in Group Theory}. Springer, 1987.

\bibitem{How79}
Edward Howorka.
\newblock On metric properties of certain clique graphs.
\newblock {\em Journal of Combinatorial Theory, Series B}, 27(1):67 -- 74,
  1979.

\bibitem{JL02}
EA~Jonckheere and P~Lohsoonthorn.
\newblock A hyperbolic geometry approach to multi-path routing.
\newblock In {\em Proceedings of the 10th Mediterranean Conference on Control
  and Automation (MED 2002)}, pages 9--10, 2002.

\bibitem{JL04}
Edmond Jonckheere and Poonsuk Lohsoonthorn.
\newblock Geometry of network security.
\newblock In {\em American Control Conference, 2004. Proceedings of the 2004},
  volume~2, pages 976--981. IEEE, 2004.

\bibitem{AJL11}
Edmond~A. Jonckheere, Poonsuk Lohsoonthorn, and Fariba Ariaei.
\newblock Scaled {G}romov four-point condition for network graph curvature
  computation.
\newblock {\em Internet Mathematics}, 7(3):137--177, 2011.

\bibitem{BJL08}
Edmond~A. Jonckheere, Poonsuk Lohsoonthorn, and Francis Bonahon.
\newblock Scaled {G}romov hyperbolic graphs.
\newblock {\em Journal of Graph Theory}, 57(2):157--180, 2008.

\bibitem{Kle07}
Robert Kleinberg.
\newblock Geographic routing using hyperbolic space.
\newblock In {\em INFOCOM 2007. 26th IEEE International Conference on Computer
  Communications. IEEE}, pages 1902--1909. IEEE, 2007.

\bibitem{HKL10}
Jure Leskovec, Daniel Huttenlocher, and Jon Kleinberg.
\newblock Signed networks in social media.
\newblock In {\em Proc. of the 28th intern. conf. on human factors in computing
  systems}, pages 1361--1370. ACM, 2010.

\bibitem{FKL07}
Jure Leskovec, Jon Kleinberg, and Christos Faloutsos.
\newblock Graph evolution: Densification and shrinking diameters.
\newblock {\em ACM Transactions on Knowledge Discovery from Data (TKDD)},
  1(1):2, 2007.

\bibitem{DLLM09}
Jure Leskovec, Kevin~J Lang, Anirban Dasgupta, and Michael~W Mahoney.
\newblock Community structure in large networks: natural cluster sizes and the
  absence of large well-defined clusters.
\newblock {\em Internet Mathematics}, 6(1):29--123, 2009.

\bibitem{Loh03}
Poonsuk Lohsoonthorn.
\newblock {\em Hyperbolic geometry of networks}.
\newblock PhD thesis, University of Southern California, 2003.

\bibitem{NS11}
Onuttom Narayan and Iraj Saniee.
\newblock Large-scale curvature of networks.
\newblock {\em Physical Review E}, 84(6):066108, 2011.
\newblock 2009 arXiv.org version \url{http://arxiv.org/abs/0907.1478}.

\bibitem{NST12}
Onuttom Narayan, Iraj Saniee, and Gabriel~H. Tucci.
\newblock Lack of spectral gap and hyperbolicity in asymptotic
  {E}rd\"os-{R}enyi sparse random graphs.
\newblock In {\em 5th Intern. Symp. Comm. Control and Signal Proc. (ISCCSP)},
  pages 1--4, May 2012.
\newblock Full arXiv.org version \url{http://arxiv.org/pdf/1009.5700.pdf}.

\bibitem{N03}
Mark E.~J. Newman.
\newblock The structure and function of complex networks.
\newblock {\em SIAM Review}, 45(2):167--256, 2003.

\bibitem{FIR02}
Matei Ripeanu, Adriana Iamnitchi, and Ian Foster.
\newblock Mapping the {G}nutella network.
\newblock {\em IEEE Internet Computing}, 6(1):50--57, January 2002.

\bibitem{Makse}
H.D. Rozenfeld, C.~Song, and H.A Makse.
\newblock Small world to fractal transition in complex networks: a
  renormalization group approach.
\newblock {\em Physical Review Letters}, 104(2):025701, 2010.

\bibitem{ST08}
Yuval Shavitt and Tomer Tankel.
\newblock Hyperbolic embedding of internet graph for distance estimation and
  overlay construction.
\newblock {\em Networking, IEEE/ACM Transactions on}, 16(1):25--36, 2008.

\bibitem{MSW02}
Neil Spring, Ratul Mahajan, and David Wetherall.
\newblock Measuring {I}{S}{P} topologies with {R}ocketfuel.
\newblock {\em ACM SIGCOMM Computer Communication Review}, 32(4):133--145,
  2002.

\bibitem{UKBM11}
Johan Ugander, Brian Karrer, Lars Backstrom, and Cameron Marlow.
\newblock The anatomy of the {F}acebook social graph.
\newblock {\em arXiv preprint arXiv:1111.4503}, 2011.

\bibitem{CGMV09}
Bimal Viswanath, Alan Mislove, Meeyoung Cha, and Krishna~P. Gummadi.
\newblock On the evolution of user interaction in {F}acebook.
\newblock In {\em Proceedings of the 2nd ACM SIGCOMM Workshop on Social
  Networks (WOSN'09)}, August 2009.

\bibitem{WS98}
Duncan Watts and Steven Strogatz.
\newblock Collective dynamics of small-world networks.
\newblock {\em Nature}, 393:440--442, 1998.

\bibitem{PSWZ12}
Christo Wilson, Alessandra Sala, Krishna P.~N. Puttaswamy, and Ben~Y. Zhao.
\newblock Beyond social graphs: user interactions in online social networks and
  their implications.
\newblock {\em ACM Trans. Web}, pages 17:1--17:31, 2012.

\bibitem{SALA11}
Xiaohan~Zhao Zhao, Alessandra Sala, Haitao Zheng, and Ben~Y. Zhao.
\newblock Efficient shortest paths on massive social graphs.
\newblock In {\em Proc. of 7th Intern. Conf. on Collab. Comput.}, Oct. 2011.

\end{thebibliography}
\bibliographystyle{plain}

\end{document}